\documentstyle [fleqn,12pt] {article}

\begin{document}
\begin{center}
{\Large \bf Uncertainties in Coupling Constant Unification}\\
\vspace{6ex}
Paul Langacker and Nir Polonsky\\
 {\it Department of Physics, University of Pennsylvania\\
 Philadelphia, Pennsylvania, 19104, USA }\\
 October 14, 1992, UPR-0513T\\

\vspace{6ex}
ABSTRACT\\
\vspace{3ex}
\end{center}
The status of coupling constant unification in the standard
model and its supersymmetric extension are discussed. Uncertainties
associated with the input coupling constants, $m_{t}$,
threshold corrections
at the low and high scales, and possible nonrenormalizable operators
are parametrized and estimated. A simple parametrization of a general
supersymmetric new particle spectrum is given. It is shown that an
effective scale $M_{SUSY}$ can be defined, but for a realistic spectrum
it may differ considerably from the typical new particle masses. The
implications of the lower (higher) values of $\alpha_{s}(M_{Z})$
suggested by low-energy ($Z$-pole) experiments are discussed.
\vspace{8ex}
\newpage

\newcommand{\gev}{\mbox{GeV}}
\newcommand{\sgev}{\mbox{ \tiny GeV}}
\newcommand{\mz}{\mbox{$ M_{Z}$} }
\newcommand{\mzns}{\mbox{$ M_{Z}$}}
\newcommand{\mx}{\mbox{$ M_{\zeta}$} }
\newcommand{\mth}{\mbox{$ M_{threshold}$} }
\newcommand{\mb}{\mbox{$ M_{boundary}$} }
\newcommand{\mbns}{\mbox{$ M_{boundary}$}}
\newcommand{\mg}{\mbox{$ M_{G}$} }
\newcommand{\mgns}{\mbox{$ M_{G}$}}
\newcommand{\mt}{\mbox{$ m_{t}$} }
\newcommand{\mtns}{\mbox{$ m_{t}$}}
\newcommand{\mbt}{\mbox{$ m_{b}$} }
\newcommand{\ooo}{\mbox{$ 138$}GeV }
\newcommand{\mttg}{\mbox{138 GeV}}
\newcommand{\mttt}{\mbox{$ \frac{m_t}{138{\sgev}} $} }
\newcommand{\mtgev}{\mbox{$ \frac{m_t}{\sgev} $} }
\newcommand{\mtttns}{\mbox{$ \frac{m_t}{138{\sgev}} $}}
\newcommand{\mh}{\mbox{$ m_{h^{0}}$} }
\newcommand{\mgg}{\mbox{$ M_{\frac{1}{2}}$} }
\newcommand{\mggns}{\mbox{$ M_{\frac{1}{2}}$}}
\newcommand{\mss}{\mbox{$ m_{0}$} }
\newcommand{\mssns}{\mbox{$ m_{0}$}}
\newcommand{\mmx}{\mbox{$ \mu_{mixing}$} }
\newcommand{\mmxns}{\mbox{$ \mu_{mixing}$}}
\newcommand{\mii}{\mbox{$ M_{i}$} }
\newcommand{\miia}{\mbox{$ M_{1}$} }
\newcommand{\miib}{\mbox{$ M_{2}$} }
\newcommand{\miic}{\mbox{$ M_{3}$} }
\newcommand{\miins}{\mbox{$ M_{i}$}}
\newcommand{\miians}{\mbox{$ M_{1}$}}
\newcommand{\miibns}{\mbox{$ M_{2}$}}
\newcommand{\miicns}{\mbox{$ M_{3}$}}
\newcommand{\miip}{\mbox{$ M_{i}^{'}$} }
\newcommand{\miipa}{\mbox{$ M_{1}^{'}$} }
\newcommand{\miipb}{\mbox{$ M_{2}^{'}$} }
\newcommand{\miipc}{\mbox{$ M_{3}^{'}$} }
\newcommand{\miipns}{\mbox{$ M_{i}^{'}$}}
\newcommand{\miipans}{\mbox{$ M_{1}^{'}$}}
\newcommand{\miipbns}{\mbox{$ M_{2}^{'}$}}
\newcommand{\miipcns}{\mbox{$ M_{3}^{'}$}}
\newcommand{\miza}{\mbox{$ \ln({\frac{M_{1}}{M_{Z}}})$} }
\newcommand{\mizb}{\mbox{$ \ln({\frac{M_{2}}{M_{Z}}})$} }
\newcommand{\mizc}{\mbox{$ \ln({\frac{M_{3}}{M_{Z}}})$} }
\newcommand{\mizi}{\mbox{$ \ln({\frac{M_{i}}{M_{Z}}})$} }
\newcommand{\miga}{\mbox{$ \ln({\frac{M_{1}^{'}}{M_{G}}})$} }
\newcommand{\migb}{\mbox{$ \ln({\frac{M_{2}^{'}}{M_{G}}})$} }
\newcommand{\migc}{\mbox{$ \ln({\frac{M_{3}^{'}}{M_{G}}})$} }
\newcommand{\migi}{\mbox{$ \ln({\frac{M_{i}^{'}}{M_{G}}})$} }
\newcommand{\mxz}{\mbox{$ \ln({\frac{M_{\zeta}}{M_{Z}}})$} }
\newcommand{\mxg}{\mbox{$ \ln({\frac{M_{\zeta}}{M_{G}}})$} }
\newcommand{\mtz}{\mbox{$ \ln({\frac{m_{t}}{M_{Z}}})$} }
\newcommand{\mtm}{\mbox{$ \ln({\frac{m_{t}}{138 \sgev}})$} }
\newcommand{\mmz}{\mbox{$ \ln({\frac{138 \sgev} {M_{Z}}})$} }
\newcommand{\mxb}{\mbox{$ \ln({\frac{M_{\zeta}}{M_{boundary}}})$} }
\newcommand{\mv}{\mbox{$ M_{V}$} }
\newcommand{\mvv}{\mbox{$ M_{24}$} }
\newcommand{\mvvv}{\mbox{$ M_{5} $} }
\newcommand{\mvns}{\mbox{$ M_{V}$}}
\newcommand{\mvvns}{\mbox{$ M_{24}$}}
\newcommand{\mvvvns}{\mbox{$ M_{5} $}}
\newcommand{\mvxa}{\mbox{$ \ln({\frac{M_{V}}{M_{G}}})$} }
\newcommand{\mvxb}{\mbox{$ \ln({\frac{M_{24}}{M_{G}}})$} }
\newcommand{\mvxc}{\mbox{$ \ln({\frac{M_{5}}{M_{G}}})$} }
\newcommand{\msus}{\mbox{$ M_{SUSY} $} }
\newcommand{\msz}{\mbox{$ \ln({\frac{M_{SUSY}}{M_{Z}}})$} }
\newcommand{\mheavy}{\mbox{$ M_{heavy} $} }
\newcommand{\mhg}{\mbox{$ \ln({\frac{M_{heavy}}{M_{G}}})$} }
\newcommand{\msusns}{\mbox{$ M_{SUSY} $}}
\newcommand{\mszns}{\mbox{$ \ln({\frac{M_{SUSY}}{M_{Z}}})$}}
\newcommand{\mheavyns}{\mbox{$ M_{heavy} $}}
\newcommand{\mhgns}{\mbox{$ \ln({\frac{M_{heavy}}{M_{G}}})$}}
\newcommand{\mpk}{\mbox{$ M_{planck} $} }
\newcommand{\mpp}{\mbox{$ M_{planck}^{-1} $} }
\newcommand{\mppns}{\mbox{$ M_{planck}^{-1} $}}
\newcommand{\ai}{\mbox{$ \frac{1}{\alpha_{i}(M_{Z})}$} }
\newcommand{\htt}{\mbox{$ h_{t}$} }
\newcommand{\hbb}{\mbox{$ h_{b}$} }
\newcommand{\httns}{\mbox{$ h_{t}$}}
\newcommand{\hbbns}{\mbox{$ h_{b}$}}
\newcommand{\topbot} {\mbox{$ {\frac {\mt\hbb}{\mbt\htt}}$}}
\newcommand{\aig}{\mbox{$ \frac{1}{\alpha_{i}(M_{G})}$} }
\newcommand{\alp}{\mbox{$\frac{1}{\alpha(M_{Z})}$} }
\newcommand{\as}{\mbox{$ \frac{1}{\alpha_{s}(M_{Z})}$} }
\newcommand{\ag}{\mbox{$ \frac{1}{\alpha_{G}}$} }
\newcommand{\agns}{\mbox{$\frac{1}{\alpha_{G}}$}}
\newcommand{\ali}{\mbox{$  \alpha_{i}(M_{Z})$} }
\newcommand{\alj}{\mbox{$  \alpha_{j}(M_{Z})$} }
\newcommand{\alig}{\mbox{$ \alpha_{i}(M_{G})$} }
\newcommand{\aljg}{\mbox{$ \alpha_{j}(M_{G})$} }
\newcommand{\allp}{\mbox{$ \alpha(M_{Z})$} }
\newcommand{\allpns}{\mbox{$ \alpha(M_{Z})$}}
\newcommand{\als}{\mbox{$ \alpha_{s}(M_{Z})$} }
\newcommand{\alsns}{\mbox{$ \alpha_{s}(M_{Z})$}}
\newcommand{\alg}{\mbox{$ \alpha_{G}$} }
\newcommand{\algns}{\mbox{$ \alpha_{G}$}}
\newcommand{\th}{\mbox{$  \theta_{i}$} }
\newcommand{\thns}{\mbox{$  \theta_{i}$}}
\newcommand{\del}{\mbox{$  \Delta_{i}$} }
\newcommand{\delns}{\mbox{$  \Delta_{i}$}}
\newcommand{\dell}{\mbox{$  \Delta $} }
\newcommand{\delc}{\mbox{$  \Delta_{i}^{conversion}$} }
\newcommand{\deln}{\mbox{$  \Delta_{i}^{NRO}$} }
\newcommand{\delt}{\mbox{$  \Delta_{i}^{top}$} }
\newcommand{\deltns}{\mbox{$  \Delta_{i}^{top}$}}
\newcommand{\dely}{\mbox{$  \Delta_{i}^{Yukawa}$} }
\newcommand{\delyns}{\mbox{$  \Delta_{i}^{Yukawa}$}}
\newcommand{\btf}{\mbox{$  \beta$} function }
\newcommand{\bi}{\mbox{$ b_{i}$} }
\newcommand{\bins}{\mbox{$ b_{i}$}}
\newcommand{\bij}{\mbox{$ b_{ij}$} }
\newcommand{\bijns}{\mbox{$ b_{ij}$}}
\newcommand{\bit}{\mbox{$ b_{i;top}$} }
\newcommand{\bitns}{\mbox{$ b_{i;top}$}}
\newcommand{\sth}{\mbox{$ s^{2}(M_{Z})$} }
\newcommand{\sthns}{\mbox{$ s^{2}(M_{Z})$}}
\newcommand{\stho}{\mbox{$ s^{2}_{0}(M_{Z})$} }
\newcommand{\sthons}{\mbox{$ s^{2}_{0}(M_{Z})$}}
\newcommand{\sths}{\mbox{$ s^{2}$}}
\newcommand{\sthso}{\mbox{$ s^{2}_{0}$}}
\newcommand{\ms}{\mbox{$ \overline{MS}$} }
\newcommand{\dr}{\mbox{$ \overline{DR}$} }
\newcommand{\msns}{\mbox{$ \overline{MS}$}}
\newcommand{\drns}{\mbox{$ \overline{DR}$}}
\newcommand{\gi}{\mbox{$ G_{i}$} }
\newcommand{\ci}{\mbox{$ C_{2}(\gi)$} }
\newcommand{\jx}{\mbox{$ J_{\zeta}$} }
\newcommand{\cjx}{\mbox{$ C^{J_{\zeta}}$} }
\newcommand{\cjv}{\mbox{$ C^{1}_{\overline{MS}}$} }
\newcommand{\cjf}{\mbox{$ C^{\frac{1}{2}}_{\overline{MS}}$} }
\newcommand{\cjs}{\mbox{$ C^{0}_{\overline{MS}}$} }
\newcommand{\cdr}{\mbox{$ C^{J}_{\overline{DR}}$} }
\newcommand{\sul}{\mbox{$ SU(2)_{L}$} }
\newcommand{\suc}{\mbox{$ SU(3)_{c}$} }
\newcommand{\uy}{\mbox{$ U(1)_{\frac{Y}{2}}$} }
\newcommand{\sss}{\mbox{$ S = \suc \otimes $}}
\newcommand{\ssss}{\mbox{$ \sul \otimes \uy $}}
\newcommand{\sug}{\mbox{$ SU(5)$} }
\newcommand{\sugns}{\mbox{$ SU(5)$}}
\newcommand{\sun}{\mbox{$ SU(N)$} }
\newcommand{\uu}{\mbox{$ U(1)$} }
\newcommand{\uuns}{\mbox{$ U(1)$}}
\newcommand{\nn}{\mbox{$ N $} }
\newcommand{\gb}{\mbox{$ \underline{24}$} }
\newcommand{\ch}{\mbox{$ \underline{5}$} }
\newcommand{\ei}{\mbox{$ \epsilon_{i}$} }
\newcommand{\ki}{\mbox{$ k_{i}$} }
\newcommand{\etaa}{\mbox{$ \eta $} }
\newcommand{\xx}{\mbox{$ \zeta $} }
\newcommand{\xxns}{\mbox{$ \zeta $}}
\newcommand{\di}{\mbox{$ D$} }
\newcommand{\dins}{\mbox{$ D$}}
\newcommand{\zi}{\mbox{$ Z$} }
\newcommand{\zins}{\mbox{$ Z$}}
\newcommand{\ti}{\mbox{$ t$} }
\newcommand{\tins}{\mbox{$ t$}}
\newcommand{\minus}{\mbox{$\--$}}
\newcommand{\for}{\mbox{  for $i$ {$=1,2,3,$}}}
\newcommand{\forr}{\mbox{  for $i$ {$=1,2,3.$}}}
\newcommand{\aaa}{{(\sl a)} }
\newcommand{\bbb}{{(\sl b)} }
\newcommand{\aaans}{{(\sl a)}}
\newcommand{\bbbns}{{(\sl b)}}
\newcommand{\aco}{\mbox{$\acute{\mbox{o}}$}}
\newcommand{\ddo}{\mbox{$\ddot{\mbox{o}}$}}
\newcommand{\du}{\mbox{$\ddot{\mbox{u}}$}}
\newcommand{\chc}{\mbox{$\check{\mbox{c}}$}}
\newcommand{\acc}{\mbox{$\acute{\mbox{c}}$}}
\newcommand{\aca}{\mbox{$\acute{\mbox{a}}$}}
\newcommand{\tn}{\mbox{$\tilde{\mbox{n}}$}}
\newcommand{\free}{\mbox{ free}}
%%%%%
\newcommand{\dela}{\mbox{$  \Delta_{1}$}}
\newcommand{\delb}{\mbox{$  \Delta_{2}$}}
\newcommand{\deld}{\mbox{$  \Delta_{3}$}}
\newcommand{\delv}{\mbox{$  \Delta_{i}^{V}$}}
\newcommand{\delvv}{\mbox{$  \Delta_{i}^{24}$}}
\newcommand{\delvvv}{\mbox{$  \Delta_{i}^{5}$}}
\newcommand{\dels}{\mbox{$  \Delta_{i}^{SUSY}$}}
\newcommand{\tha}{\mbox{$  \theta_{1}$}}
\newcommand{\thb}{\mbox{$  \theta_{2}$}}
\newcommand{\thc}{\mbox{$  \theta_{3}$}}
\newcommand{\mia}{\mbox{$ M_{1}$} }
\newcommand{\mib}{\mbox{$ M_{2}$} }
\newcommand{\mic}{\mbox{$ M_{3}$} }
\newcommand{\ova}{\mbox{$ \frac{41}{10} $}}
\newcommand{\ovb}{\mbox{$ -{\frac{19}{6}} $}}
\newcommand{\ovc}{\mbox{$ \frac{66}{10} $}}
\newcommand{\ddd}{\mbox{$ D \equiv 5b_{1} + 3b_{2} - 8b_{3}$}}
\newcommand{\ba}{\mbox{$ b_{1} $}}
\newcommand{\bb}{\mbox{$ b_{2} $}}
\newcommand{\bc}{\mbox{$ b_{3} $}}

 %1111
 \section{Introduction}
 Implications of precision \zins-pole, $W$ mass, and neutral current
 data for the standard model were considered previously
 in ref. \cite{pgl1}.
 Constraints on  the top mass were derived, and the value of weak angle
 at the \zins-pole, $\sin^{2}\theta_{W}(\mz)$, was extracted from
 the data. It was further shown that
 within the supersymmetric \sug grand unified theory (GUT)
 \cite{su5} -- \cite{rossrev}
 the two-loop prediction \cite{gqw}
 of the weak angle agrees well with the value
 extracted from the data; that
 the standard model couplings meet at a point (within the
 \als uncertainty) when extrapolated to high energy;
 and that the scale at which they meet is
 high enough to prevent a too fast proton
 decay rate via vector boson exchange.
 On the other hand, when assuming the ordinary \sug GUT the
 standard model couplings, $\alpha_{1},\,\alpha_{2},$ and  $\alpha_{3}$,
 do not meet, and the predicted proton decay rate is much too rapid.
 Similar observations were made by other groups \cite{others}.
 Here and below, we denote the coupling of the group $G_{i}$ by
 $\alpha_{i}$, where $G_{i} = U(1)_{\frac{Y}{2}}, \, SU(2)_{L}, \,
 SU(3)_{c}$ for $i$ = 1, 2, 3, respectively, and $\alpha_{1}$
 is further normalized as required.
 All of the couplings, as well as the weak angle, are defined in this
 paper in the modified minimal substraction scheme (\msns)
 \cite{ms,mss}
 unless otherwise specified. The \ms weak angle will be denoted
 below by $s^{2}$.

 The above observations are true for a whole class of
 GUT's which break to the standard model group in one step,
 and which predict
 a ``grand desert'' between the weak (low-)
 and the  grand unification (high-) scales (one-step GUT's).
 In particular, they hold for larger groups such as $SO(10)$ and
 $E_{6}$ which have the same relative normalization of the $G_{i}$
 generators, provided there are no additional matter (super)multiplets
 that are split into light and heavy components.
 However, the \sug model has the
 minimal gauge group and, in the simplest version,
 a minimal matter content, and is therefore useful for illustration.
 One should note that high-scale thresholds can modify the
 predictions, and thus in principle distinguish different one-step GUT's.
 If a ``grand desert'' indeed exists,
 and, furthermore, supersymmetry is established and characterized at
 future colliders, we may eventually be able to use coupling constant
 unification to probe the physics near the unification and Planck scales.

 We dedicate most of this paper to a more thorough discussion of one-step
 GUT's. Let us mention, however, that one could also
 fit the data to a model in which intermediate scales are
 introduced. In ref. \cite{pgl1} left-right models
 (derived from non-supersymmetric $SO(10)$ GUT's)
 \cite{pgl2,rossbook,pgl5} were considered, and it was found that
 models with an intermediate scale $M_{R} \approx 10^{10}$ GeV
 for the breaking of the right-handed $SU(2)_{R}$
 are consistent with the data. (The  supersymmetric version of the model
 requires that $M_{R}$ is close to the unification scale \cite{pgl1}.)
 A more recent discussion of $SO(10)$ models is given in ref. \cite{moh}.
 Models involving ad hoc new matter
 multiplets split into light and superheavy components
 were also considered \cite{inter}.
 Such models lose most of the predictive power of the ordinary or
 supersymmetric grand desert theories, because either the intermediate
 scales or the quantum numbers of the new multiplets are chosen to fit
 the data.
 We do not discuss such possibilties any further in this paper.

 The better standing of the supersymmetric one-step GUT's
 compared to the ordinary ones has been known for
 some time \cite{su5s2,j2,amaldi}.
 However, the much more precise coupling constant data from
 LEP \cite{lep} has shown this more strongly
 and motivated a revived interest in GUT's.
 As we will show below, with such precise inputs the predictions
 become sensitive to small
 correction terms (threshold corrections and others)
 which are often ignored.
 Recently, detailed calculations of the supersymmetry (SUSY) new
 particle (sparticle) spectrum were carried out
 \cite{ross1,nath,klnpy},
 and constraints from proton decay via dimension-five
 operators \cite{nath}, and from fine-tuning of the top mass
 \cite{ross1,klnpy}, were again considered.
 The possible equivalence of
 threshold corrections at the low- and  high- scales was pointed out
 in ref. \cite{h2}\footnote{Constraints from proton decay were ignored in
 ref. \cite{h2}, as discussed in ref. \cite{farag}.
 Barbieri and Hall's conclusion in ref. \cite{h2} is, however,
 a qualitative one, and still holds. In both ref. \cite{h2} and
 \cite{farag} the naive effective parameter \msus was used.
 Below, we show that $\msus < \mz$ is allowed when sparticle mass
 splittings are included.}.
 It was also shown that
 in SUSY GUT's with large representations, for which
 sterile neutrinos can have large masses comparable to the unification
 scale, the light neutrino masses predicted in seesaw
 models \cite{seesaw} are
 smaller than those suggested by the solar
 neutrino problem for $\nu_{e} \longrightarrow \nu_{\mu}$ oscillations
 \cite{pgl6}. The possible role of non-renormalizable operators (NRO's)
 at the high-scale for generating more suitable neutrino masses
 has been pointed out \cite{pgl4}.
 A more careful consideration of the
 model predictions, and in a way that consistently incorporates different
 correction terms that may be significant -- individually or
 cumulatively -- is now required.

 Some of the possible correction terms were considered recently in ref.
 \cite{ekn,az,yy}.
 In ref. \cite{yy} threshold corrections at the high-scale
 were discussed while the sparticle ones were treated naively. In ref.
 \cite{ekn,az} sparticle thresholds were discussed in detail
 and used to constrain the high-scale
 gaugino mass parameter. The motivation and approach here are
 different. We will suggest below an alternative way to treat the
 sparticle thresholds. We will elaborate on an observation of Ross
 and Roberts \cite{ross1} that a naive analysis, in which all
 sparticles and new Higgs particles are degenerate at a scale \msus
 \cite{pgl1,others,h2,farag,yy}, can be misleading, e.g.,
 because the average mass of the colored sparticles may be larger than
 that of the
 uncolored ones. We give a simple parametrization of the effects of an
 arbitrary sparticle spectrum and show that an effective \msus can
 always be defined. However, for realistic splittings \msus can differ
 drastically from the actual sparticle masses, and, in particular, one
 can have $\msus < \mz $ (as is suggested if \als is sufficiently large)
 even though the actual sparticle masses are much larger than \mzns.
 We will also treat the heavy
 $t$-quark threshold corrections and the $m_{t}$ contribution to the
 input parameter uncertainties consistently,
 and will consider threshold and
 NRO correction terms at the high-scale. A convenient
 parametrization
 of the high-scale threshold corrections will be suggested as well.

 Below, we will use the following (updated) input values of the
 low-scale parameters:
 \begin{equation}
 \mz = 91.187 \pm 0.007 \; \gev.
 \label{input5}
 \end{equation}
 A two-parameter fit to all $Z$, $W$, and neutral current data yields
 \begin{equation}
 \sth = 0.2324 \pm 0.0006 \; \; \; (\mt  \free)
 \label{input1}
 \end{equation}
 \begin{equation}
 {\mbox{$ \mt = 138^{+20}_{-25} \pm 5$}\; \gev},
 \label{input2}
 \end{equation}
 where the  the central values assume\footnote{ For $m_{h^{0}}$ varying
 from $50-1000$ GeV with a central value of 250 GeV, as is reasonable
 for the non-supersymmetric standard model, one obtains
 $\sth = 0.2325 \pm 0.0007$,
 {\mbox{$  \mt = 150^{+17 + 15}_{-23 - 17}$}} GeV.}
 a Higgs mass $m_{h^{0}} = \mz$. The second error in
 \mt is from allowing $m_{h^{0}}$ to vary from  50 - 150 GeV, which is
 a reasonable range for the light Higgs scalar in the minimal
 supersymmetric extension of the standard model. (The additional Higgs
  particles do not contribute significantly to the experimental
 determination of \sthns. Their contributions to the running are treated
 as threshold corrections.) Most of the uncertainty in \sth is due to
 \mt and $m_{h^{0}}$.
  It is convenient to use the more restrictive value
 \begin{equation}
 \stho = 0.2324 \pm 0.0003 \; \; \; (\mt = 138 \; \gev),
 \label{input6}
 \end{equation}
 which is obtained for the fixed values $\mt = 138$ GeV,
 $m_{h^{0}} = \mz$. The uncertainties from \mt and $m_{h^{0}}$ will be
 treated separately.

 We also have\footnote{There is a weak correlation
 between the \allp and \stho error bars, associated with the hadronic
 contribution to the running of $\alpha$. The effect is numerically
 insignificant to the discussion.}
 \cite{s1}
 \begin{equation}
 \alp = 127.9 \pm 0.1,
 \label{input3}
 \end{equation}
 which is valid for $\mt = 138$ GeV. Note that the values used here for
 \allp and \sth correspond to the definitions in \cite{s1}.
 This is not quite the canonical \ms because \mt is not decoupled, i.e.,
 it contributes to the running even below \mtns. We will correct for this
 and treat the uncertainty from \mt in the threshold corrections.

 The largest uncertainty in the input parameters is from \alsns. Some of
 the more precise determinations are shown in Table 1 and Figure 1, which
 are adopted from a recent review of Bethke and Catani \cite{alphas}
 (see also \cite{hebb}). It is seen that there is a tendency for the
 lower energy measurements to yield smaller \als than the $Z$-pole
 determinations\footnote{This has even prompted the suggestion that
 there may be a light gluino which modifies the extrapolation
 \cite{glouino}.}.
 However, all of the determinations except $R$ (which
 still has a large statistical error) have considerable theoretical
 uncertainties which could very well be underestimated, so there is no
 compelling evidence for a discrepancy. We will take\footnote{This is
 higher than the value $0.1134 \pm 0.0035$ given in the Review of
 Particle Properties \cite{rpp} due to the use of resummed QCD
 \cite{alphas} and a more conservative estimate of theoretical
 uncertainties.}
 \begin{equation}
 \als = 0.120 \pm 0.010
 \label{input4}
 \end{equation}
 as a reasonable estimate, for which we have assigned a
 fairly conservative uncertainty.

 The ability of GUT's to predict $s^{2}$ at the unification point ($s^{2}
 = \frac{3}{8}$ in \sug and similar models \cite{su5,gqw})
 historically led to using the prediction for \sth (\stho in our case)
 from \allp and \als as a test of the models.
 However,  the large uncertainty in \als leads to a large uncertainty
 in the predictions, and the different input values assumed by various
 authors have led to some confusion. We therefore find it
 more instructive to use \stho
 as an input in order to predict
 \alsns. We will consider both alternatives below.

  In this paper we discuss in detail the \sug grand unification
  of the standard model (with one Higgs doublet) (SM), and of the
  minimal supersymmetric standard model (with two Higgs doublets)
  (MSSM). In section 2 we review the predictions of these models,
  where we use \alsns, and alternatively \sthons, as an input.
  In section 3 we discuss in detail different correction terms
  that may affect these predictions. We introduce three effective
  mass parameters that conveniently sum
  the threshold corrections near \mzns.
  In section 4
  we collect our results and choose reasonable ranges for the
  different correction terms. We then obtain (in the MSSM)
  the predictions
  \begin{equation}
  \stho = 0.2334 \pm 0.0025 \,\pm 0.0014 \, \pm 0.0006 \,
  ^ {+0.0013}_{-0.0005} \,  \pm 0.0016
  \label{output1}
  \end{equation}
  \begin{equation}
  \als = 0.125  \pm 0.001 \, \pm 0.005 \, \pm 0.002 \,
          ^{+0.005}_{-0.002}  \, \pm 0.006,
  \label{output2}
  \end{equation}
  where the central values are for $\mt = 138$ GeV and $M_{SUSY} =
  m_{h^{0}} = M_{Z}$.
  The first uncertainty in (\ref{output1}) (in (\ref{output2}))
  is due to the \als  (\sthons)  and \allp error
  bars, and the other uncertainties in  both (\ref{output1}) and
  (\ref{output2}) are due to sparticle thresholds, \mt and $m_{h^{0}}$,
  thresholds
  at the high-scale, and NRO's at the high-scale, respectively.
  The uncertainties quoted here refer to our choice of ranges for the
  different correction terms, and should be taken as such (i.e., as
  order of magnitude estimations rather than rigorous ranges).
  Note that the different theoretical uncertainties are comparable to the
  \als error bar in (\ref{input4}) and to the corresponding
  uncertainty in (\ref{output1}). The combined
  theoretical uncertainty is determined by an interplay among the
  different terms, most of which can have either sign. (If the
  high-scale thresholds are not constrained as in the minimal model (see
  below),  then none of the uncertainties has a fixed sign.)
  When added in quadrature, the above theoretical uncertainties yield a
  $+ 0.0026 - 0.0023$ ($+0.010 - 0.008$) combined uncertainty in the
  \stho (\alsns) prediction. The predicted \als is compared
  with the data in Figure 1, while the \sth prediction is shown in
  Figure 2. The extrapolated coupling constants are shown in Figure 3.
  Corresponding predictions for the unification scale and the coupling
  at that scale are given in section 4.
  In all cases, it is seen that the MSSM (but not the SM) is in agreement
  with the prediction of unification.

  The prediction for \als is in good agreement with the value observed
  at the $Z$-pole, and the larger $Z$-pole value for \als predicts a
  smaller \sthons, in agreement with observation. The somewhat lower \als
  values suggested by low-energy experiments could be accomodated
  by $\msus > \mz$, $\mt < 138$ GeV, or the introduction of NRO's.
  Also, in the simplest SUSY-\sug the high-scale thresholds increase the
  predicted \als when constraints from proton decay are included.
  However, simple extensions, e.g.,
  replacing R-parity with baryon-parity \cite{ross2},
  or the introduction of additional matter supermultiplets
  at the high-scale, would allow smaller \alsns.
  We discuss the high-scale thresholds in
  more general terms in section 5, where we introduce
  effective parameters, similar to those introduced for the sparticles
  in section 3.
  Throughout this paper we display the various expressions
  in a transparent form, which enables one to generalize our
  discussion and to use the results elsewhere. We summarize
  our conclusions in section 6.

  %2222
  \newpage
  \section{One- and Two- Loop Predictions}
  When solving for the running of the couplings in any GUT scenario
  with no intermediate scale, we can reduce the problem to one of
  a ``grand desert'' and account for all thresholds near the desert
  boundaries by properly defining correction terms.
  If one uses a two-loop \btf for the running, then one-loop threshold
  corrections \cite{w,h1,thresh} usually suffice.
  The normalized couplings are then \cite{h1,j2}
  \begin{equation}
  \ai =  \ag + b_{i}t + \th - \del, \; \for
  \label{rge}
  \end{equation}
  where {\mbox{$ \ti \equiv {\frac{1}{(2\pi)}}
  \ln({\frac{M_{G}}{M_{Z}}})$}},
  \mg is the grand unification scale (which serves as the high-scale
  boundary of the desert), and \alg
  is the coupling at that point.
  {\mbox{$ \th \equiv {\frac{1}{(4\pi)}} \sum_{j = 1}^{3}{\frac{b_{ij}}
  {b_{j}}} \ln({\frac{\alpha_{j}(M_{G})}{\alpha_{j}(M_{Z})}}) $}}
  are the two-loop terms, and \bi (\bijns)
  are the one- (two-) loop
  \btf
  coefficients,
  \begin {equation}
  \mu{\frac{d\alpha_{i}}{d\mu}} = {\frac{\bi}{(2\pi)}}\alpha_{i}^{2}
  + \sum_{j = 1}^{3}{\frac{\bij}{(8\pi^{2})}}\alpha_{i}^{2}\alpha_{j},
  \label {beta}
  \end{equation}
  which can be calculated using, for example, ref. \cite{j1,mv}.

  \del are  threshold and other corrections, which should be calculated
  to a precision consistent with the \thns.
  Our ignorance of their exact values
  suggests that they should be reasonably parametrized
  and estimated within a given model, and then translated into
  theoretical uncertainties on any predictions.  This will be carried out
  in the following sections.
  We will also show that for reasonable masses for the sparticles,
  the MSSM can be treated as a two-scale model with all mass effects
  included in the threshold corrections.

  At the \zi threshold (which serves as the low-scale boundary of the
  desert), we have
  {\mbox{$\ai = {\frac{3}{5}}{\frac{(1 - s^{2}(M_{Z}))}{\alpha(M_{Z})}},
  {\frac{s^{2}(M_{Z})}{\alpha(M_{Z})}},
   \as $} }
  for  $i$ = 1, 2, 3, respectively. \sth (which we replace with \sthons),
  \allpns, \als are the three low-scale (\ms) parameters
  defined previously, evaluated at the $Z$-pole.
  By taking linear combinations of (\ref{rge}) one obtains explicit
  expressions for the two high-scale parameters \ti and \algns, and for
  one low-scale parameter, in terms of the other two,
  the \btf one-loop coefficients and the two-loop and correction terms.

  The two-loop terms can be rewritten using the lowest order solution
  for the couplings \cite{h1,j2},
  \begin{equation}
  \th = {\frac{1}{(4\pi)}} \sum_{j = 1}^{3} {\frac{\bij}{b_{j}}}
  \ln(1 + b_{j} \alg t),
  \label{theta}
  \end{equation}
  where the  one-loop expressions for \alg and $t$ are to be
  substituted\footnote{In practice we will
  use the full two-loop values for  \ti and \alg
  in \thns, solving iteratively. The difference between the
  two procedures is of higher order.}.
  For a given model
  one can then predict \tins, \alg and either \stho -- which
  we will refer to as
  case \aaa -- or \als -- which we will refer to as case \bbb -- in terms
  of \allp and either \als or \sthons.

  We list in Tables 2a-b the general expressions for \tins, \algns, and
  \sth (\alsns\footnote{ We use a Taylor expansion to
  convert the prediction of \as to an expression
  for \alsns. In Table 2 we give the zero (one-loop) and first (two-loop)
  order terms in the expansion. This gives $\sim 99.2\% $ accuracy.
  We will include the second order term when evaluating \alsns. }), where
  we define a linear combination of the one-loop \btf
  coefficients, {\mbox{$ D \equiv 5b_{1} + 3b_{2} - 8b_{3} $}}.
  The correction term for each expression is of the same form as the
  two-loop term, only with \th replaced by \minus\delns.
  Note that we have exactly the same expressions when replacing \sth
  with \sthons, except that the $t$-quark and the Higgs particle
  contributions to the correction terms \del are different.
  For the two models studied in this paper, the SM and the MSSM,
  the $\beta$ functions
  can be found in ref. \cite{j2}, where the dependence on the
  number of fermion families and Higgs doublets is explicitly given.
  For completeness we give
  \bins, \bij and \di for the SM (MSSM) with
  our choice of three families and one (two) Higgs doublet(s) in Table 3.
  Then, using Tables 2, 3  and the input parameters, we can
  calculate the two-loop terms for each case. These are
  listed in Tables 4a-b, where we also compare the \th values calculated
  using the one-loop \ti and \algns, and those calculated
  iteratively.
  For different values of the low-scale input parameters the two-loop
  terms should be recalculated, though for
  a small change the difference is negligible.

  In Tables 5a-b we give the predictions corresponding to our central
  values of the input parameters, but not including any correction terms.
  One can clearly see that the MSSM is
  consistent with these values (see also Figures 1-3).
  Cases \aaa and \bbb in the MSSM are
  consistent with each other at the two-loop order. Also the
  prediction of \ti in that model is large
  enough to prevent an observed proton decay via a heavy vector boson
  exchange \cite{proton}.
  The value of \ti corresponds to $\mg \sim 2.5\times10^{16}$ GeV,
  so that $\tau_{p \rightarrow e^{+}\pi^{0}} \sim \mg^{4}
  \sim 3\times10^{38 \pm 1}$ yr, much larger than the experimental
  lower limit \cite{proton} of $10^{33}$ yr.
  In the SM the inconsistency
  between cases \aaa and \bbb implies that SM
  unification is inconsistent with the present values of the input
  parameters (see also Figures 1-3). Also, the SM prediction of
  \ti is inconsistent with proton decay limits\footnote{In the SM case
  one has approximately
  $\tau_{p \rightarrow e^{+}\pi^{0}}$(yr) $\sim 10^{31 \pm 1}\left(
  {\frac{M_{G}}{4.6 \times 10^{14} \; \gev}}\right)^{4}$,
  so that $\tau > 10^{33}$ yr
  corresponds to $M_{G} > 10^{15}$ GeV or $t > 4.8$. In the MSSM the
  $e^{+}\pi^{0}$ rate is suppressed both by $M_{G}^{-4}$ and also by
  additional factor of $ \sim \frac{1}{3} $ due to the smaller \algns.}
  in either case \aaa or \bbbns.
  For case \aaa (case \bbbns) one predicts the unacceptable values
  $\mg \sim 4.6\times10^{14}$ GeV and
  $\tau_{p \rightarrow e^{+}\pi^{0}} \sim 10^{31 \pm 1}$ yr
  ($ 8.5\times10^{12}$ GeV and $ 10^{24 \pm 1}$ yr).

  The above failures of the SM
  cannot be resolved by adding either more light Higgs doublets
  or additional fermion families.
  As is well known, additional fermion families represent complete
  GUT multiplets which affect all the $b_{i}$'s equally. Hence,
  the \alsns, \stho and $t$ predictions are only modified at the two-loop
  level. (\alg is affected at one-loop.)
  On the other hand, extra Higgs families are part of partial
  GUT multiplets which affect the predictions at one-loop.
  When adding $\Delta n_{H}$ Higgs doublets in case \aaans,
  the \stho prediction increases, but \ti decreases,
  increasing the proton decay rate. For $\Delta n_{H} = 6$ one has
  $\mg \sim 4\times 10^{13}$ GeV
  and $\tau_{p} \sim 6 \times 10^{26}$ yr.
  In case \bbbns, \als increases with $n_{H}$, but eventually changes
  sign, and adding enough Higgs doublets so that \ti
  has an acceptable value drives \als negative. In the
  MSSM, extra Higgs supermultiplets will destroy the successful
  predictions for \stho and \alsns.
  For completeness we display the changes
  in the predictions for additional fermion family and Higgs
  (super)multiplets in Tables 6a-b.

  %3333
  \newpage
  \section{A Formal Discussion of The Correction Terms}
  This section will be devoted to the correction terms \delns,
  \begin{eqnarray}
  \del = \delc + \sum_{boundary} \sum_{\zeta}
  {\frac{b_{i}^{\zeta}}{(2\pi)}}(\mxb  - C^{J_{\zeta}}) & &  \nonumber \\
   + \delt + \dely +\deln. & &
  \label{deli}
  \end{eqnarray}

  The first term is a constant which depends only on the gauge
  group \gi \cite{ant},
  \begin{equation}
  \delc \equiv -{\frac{\ci}{(12\pi)}},
  \label{con}
  \end{equation}
  where \ci is the quadratic casimir operator
  for the adjoint representation,
  \ci = \nn (0) for \gi = \sun (\uuns).
  \delc results from the need to use the dimenional-reduction (\drns)
  scheme
  in the MSSM, so that the algebra is kept in four dimensions \cite{j3}.
  Thus, we convert the \ms couplings above \mzns,
  \begin{equation}
  {\mbox{$ {\frac{1}{\alpha_{i_{\overline{MS}}}}}
  = {\frac{1}{\alpha_{i_{\overline{DR}}}}}
   - \delc $}.}
  \label{alpcon}
  \end{equation}
  For consistency we will use \dr
  also in the SM case, though this is not required. \alg is then given
  in its \dr definition.

  The second term sums over the one-loop threshold corrections
  \cite{h1}. {\mbox{$ b_{i}^{\zeta}$} }
  is the (decoupled) contribution of a heavy field
  \xx to the \btf coefficient \bi between \mx and \mbns. \cjx is a
  mass-independent number
  which depends on the spin \jx of \xx and on the regularization
  scheme used. In  \ms (using dimensional-regularization)
  one has\footnote{Different regularization conventions give
  {\mbox{$ \cjf = -\ln\sqrt{2}$} } \cite{w,h1}.}
  {\mbox{$ \cjv = {\frac{1}{21}},\cjf
  = \cjs = 0$} }
  \cite{h1}.
  These  are to be used
  at the low-scale boundary, while at the other boundary
  (using dimensional-reduction\footnote{When using dimensional-reduction
  the loop integrals are analytically continued away from $d = 4$
  (as for dimensional-regularization). On the other hand, the algebra of
  the fields is not continued and is kept in $d = 4$ (i.e., $g^{\mu \nu}
  g_{\mu \nu} = 4$). Therefore, no constants arise when taking the
  limit $d \rightarrow 4$ \cite{ant}.}) we have
  {\mbox{$ \cdr \equiv 0$} } \cite{ant}. (If one converts \alg back to
  its \ms definition,
  then the sum of the two conversion terms reproduces the \ms
  mass-independent term.)

  The summation in (\ref{deli}) can account for a particle threshold
  as long as two-loop terms between this threshold and the boundary
  are negligible, i.e.,
  \begin{equation}
  {\mbox{$ |\bi\alpha_{i}(\mb)\mxb| \ll 2\pi$}}.
  \label{cond}
  \end{equation}
  This allows a split of more than 3 orders of magnitude
  for all relevant cases. Thus
  (\ref{deli}) can correctly account for a reasonable sparticle spectrum.

  At the low-scale boundary we have to consider the top, Higgs, and
  sparticle thresholds. The
  \sul symmetry is broken by the top quark mass in the range
  \mzns--\mtns, questioning the validity
  of accounting for the top in the above threshold summation.
  Furthermore, the values of the input parameters and \mt are correlated
  in a complicated way.
  Similar considerations apply to the SM Higgs. We therefore omit these
  two thresholds from the summation and discuss them separately
  below.
  In the MSSM we assume a light
  SM Higgs {\mbox{$(\mh \approx \mz)$} }and a heavy decoupled doublet,
  which is included with the sparticles. Using tree-level sum rules
  \cite{ikkt} one can show that in such a limit \sul
  breaking is negligible in the Higgs sector.
  (This conclusion is still valid when radiative corrections to
  the Higgs masses \cite{higgs,higgs2} are considered.)
  We will further assume a good symmetry in the sparticle sector (i.e.,
  \sul breaking effects are typically
  {\mbox{$ < \left({\frac{m_{t}}{m_{stop}}}\right)^{2} $}}
  and are negligible for our purposes).

  In the SM we can then omit the low-scale boundary from the
  summation
  in (\ref{deli}), while in the MSSM we are left with the sparticles
  and the heavy Higgs doublet. The sparticle  and Higgs masses
  can be calculated given a small number of high-scale
  parameters -- i.e., a universal gaugino mass \mggns; a universal
  scalar mass \mssns;
  the Higgs mixing parameter \mmxns; a universal trilinear
  coupling $A$; and the top Yukawa coupling \htt
  (we omit all other Yukawa couplings) --
  by solving a set of coupled renormalization group equations (RGE's)
  \cite{ibanez,rge}\footnote{Our notation follows that of ref.
  \cite{ibanez}, aside from self-explanatory subscripts.}.
  Other mass parameters, like the universal bilinear coupling $B$,
  are related  to the parameters above by boundary conditions
  and  the constraint setting the weak breaking scale \cite{ikkt}.
  One can then solve
  the one-loop RGE's
  for a given set of parameters, and predict a specific
   sparticle spectrum \cite{ross1,nath,klnpy}.  Substituting in
  (\ref{deli})
  gives the desired correction.  However, this is a lengthy and not very
  enlightening procedure for our purpose of estimating small correction
  terms.
  We use instead a
  parametrization in terms of three low-energy effective
  parameters defined by
  \begin{equation}
  \sum_{\zeta}{\frac{b_{i}^{\zeta}}{(2\pi)}} \mxz \equiv
  {\frac{b_{i}^{MSSM} - b_{i}^{SM}}{(2\pi)}}\mizi
  \; \for
  \label{mi}
  \end{equation}
  where the summation is over the sparticles and the
  heavy Higgs doublet. The low-scale sparticle
  spectrum can be crudely parametrized in terms of the high-scale
  parameters\footnote{In the
  limit {\mbox{$\htt \rightarrow 0$} }
  this parametrization can be made exact.}
  (with a reasonable assumption about the Higgs mass)
  \cite{n1,barb,ekn}, in order to learn about the
  relationship between the high-scale parameters and \miins. One finds
  that the case {\mbox{$ \mgg \gg \mss \approx \mmx \approx \mz$} }
  ({\mbox{$ \mss \approx \mmx \gg \mgg \approx \mz$}) } corresponds
  to {\mbox{$ M_{3} \gg M_{1},M_{2}$} }
  ({\mbox{$ M_{1} \gg M_{2},M_{3}$})}\footnote{Constraints derived from
  proton decay favor $\mss \gg \mgg$ \cite{nath}. However, we then
  may have $\mmx  < \mss$. If so, $M_{3}$ and $ M_{1}$ become closer.}.
  The parameters can be split by a
  factor of a few.  As will become clear below, it is important
  to note that we do not expect to have
  {\mbox{$ M_{2} \gg M_{1}$}} and/or $M_{2} \gg M_{3}$.
  We have also calculated \mii for the realistic spectra\footnote
  { \mg and \alg of ref. \cite{ross1} differ slightly from ours due
  to different values of input parameters and a different
  calculational procedure.
  The procedure there incorporates the sparticle effects iteratively,
  and thus the \als prediction is automatically corrected
  for sparticle thresholds. Also, $\als < 0.118$ and a fine-tuning
  constraint were imposed, and
  \mt was assigned so the constraint setting the
  weak breaking scale is satisfied.
  However, constraints on the spectrum
  parameters derived from proton decay limits \cite{nath} were not
  considered. (The proton decay and fine-tuning constraints do not
  agree.) We use the spectra given
  in ref. \cite{ross1} for illustration only, and ignore minor
  inconsistencies. When small \sul breaking occurs,
  we identify a doublet threshold with that of the heavier member.
  Our results and conclusions do not depend on
  any specific choice of spectrum.}
  given in ref. \cite{ross1} (see  Table 7).
  The parameters \mii can be calculated exactly in
  any other model using (\ref{mi}), and once calculated, all correction
  terms are given below.

  The discussion so far has
  only assumed a SM gauge group, \sss \newline
  \ssss, (with the proper normalization of $\alpha_{1}$) in the desert,
  and has been independent of the GUT gauge group.
  The high-scale corrections do depend on the group.
  For definiteness we first assume that this is \sugns,
  for both the SM and the MSSM.
  A minimal choice of massive (super)multiplets at the
  high-scale is then (listing $S$ quantum numbers)
  {\mbox{$ (\overline{3},2,{\frac{5}{6}}) \oplus  c.c.$} }
  massive vector (super)multiplets\footnote{Supermultiplets
  are defined as in ref. \cite{j2}.
  A massive vector supermultiplet consists of a real massive vector, a
  Dirac spinor, and a real scalar. A Dirac (Majorana or chiral)
  supermultiplet consists of a Dirac (Majorana or Weyl) spinor and
  two (one) complex scalars.};
  {\mbox{$ (8,1,0),(1,3,0),(1,1,0)$} }
  massive real-Higgs (Majorana super)multiplets
  (embedded in a \gb of \sugns);
  and a {\mbox{$ (3,1,-{\frac{1}{3}})$} }
  complex-Higgs (Dirac super)multiplet  (embedded in a \ch of \sugns)
  \cite{pgl2,rossbook}.
  We thus introduce 3 mass parameters, \mvns, \mvvns, and
  \mvvv for the vector, real-Higgs (Majorana),
  and complex-Higgs (Dirac) (super)multiplet
  thresholds, respectively, and we assume mass degeneracy within each
  of these (super)multiplet classes.
  (We show how to generalize this in section 5.)
  We then identify
  {\mbox{$ \mg \equiv \max({\mvns,\mvvns,\mvvvns})$}},
  so that \sug is complete
  above \mgns. In the MSSM, proton decay via dimension-five operators
  constrains {\mbox{$ \mvvv \geq 10^{16}$} GeV},
  and the validity of perturbation theory in the Higgs sector  constrains
  {\mbox{$ \mvvv \leq 3\mv $}}
  \cite{nath}. This suggests {\mbox{$ \mg \equiv \mvvv $} } in the
  MSSM. Though we shall not impose this (allowing other solutions
  to the proton decay problem \cite{ross2}),
  one has to bear in mind the possible need to carefully adjust \mvvv
  in the MSSM, and the general dependence between these parameters,
  determined by the details of the Higgs sector Lagrangian.

  We now discuss the heavy top threshold. We must
  consider both the effect on the running and on the experimental
  determination of the couplings at \mzns.
  In the \ms scheme to account for {\mbox{$ \mt > \mz$} }
  one can define threshold corrections \cite{h1} to \allp and \alsns,
  i.e.,  {\mbox {$ {\frac{b_{Q}^{top}}{(2\pi)}}\mtz $} }and
  {\mbox {$ {\frac{b_{3}^{top}}{(2\pi)}}\mtz $}}, respectively, where
  {\mbox{$ b_{Q}^{top}$} }and
  {\mbox{$ b_{3}^{top}$} }are the top contributions
  to the relevant one-loop \btf
  slope. The first of these corrections is equivalent to the
  slightly nonstandard \allp
  definition of ref. \cite{s1}, which we use. Thus, for our central
  value of {\mbox{$ \mt = 138$} GeV,} our value of \allp already
  includes the top threshold correction, and we need to further correct
  \allp only for different values of \mtns. Thus
  \begin{eqnarray}
  \Delta_{\alpha}^{top} = {\frac{8}{(9\pi)}}\mtm
  & & \label{delalp} \\
  \Delta_{\alpha_{s}}^{top} =
  {\frac{1}{(3\pi)}}\mmz   + {\frac{1}{(3\pi)}}\mtm. & &
  \label{delas}
  \end{eqnarray}
  Similarly, the \mt threshold corrections are already included in the
  \sth definition of ref. \cite{s1}. However, the input value of \sth
  extracted from the data depends both quadratically and
  logarithmically on \mtns. In particular, the value $\stho = 0.2324
  \pm 0.0003$ in (\ref{input6}) is for the best fit value $\mt = \mt_{0}
  = 138$ GeV. For other \mt the corresponding \sth is
  \begin{eqnarray}
  \sths  = \sthso
  - {\frac{3G_{F}}{8\sqrt{2}\pi^{2}}}
  \sthso{\frac{1 -\sthso}{1 - 2\sthso}}
  \left((\mt)^{2} - (\mt_{0})^{2} \right),  & &
  \label{sinmt}
  \end{eqnarray}
  where {\mbox{$G_{F}$} }is the Fermi coupling,
  and we have neglected logarithmic dependences on \mtns.
  We then have {\mbox{$\sth(\mt) = \stho +\Delta_{s^{2}}^{top}$} } where
  \begin{equation}
  \Delta_{s^{2}}^{top} \approx -1.03 \times 10^{-7}\;\gev^{-2}
  \left((\mt)^{2} - (\mt_{0})^{2}\right).
  \label{delsin}
  \end{equation}
  We take the reference value of \stho in (\ref{input6}) as
  our input value for both case \aaa and case \bbbns.
  That is, \stho can be viewed
  as a convenient parametrization of the precisely known \mzns.
  The \mt dependence of the ``true" \sth in $\Delta_{s^{2}}^{top}$
  will be included together with the threhold corrections in \deltns.
  Thus
  \begin{eqnarray}
  \Delta^{top}_{1} =  {\frac{8(1- \sthns)}{(15\pi)}}
  \mtm - \frac{3}{5}\frac{\Delta_{s^{2}}^{top}}{\allp}
  & & \label{deltop1}  \\
  \Delta^{top}_{2} =  {\frac{8\sthns}{(9\pi)}}
  \mtm + \frac{\Delta_{s^{2}}^{top}}{\allp}
  &  & \label{deltop2}   \\
  \Delta^{top}_{3} = 0.04 + {\frac{1}{(3\pi)}}\mtm.
  & & \label{deltop3}
  \end{eqnarray}
  The SM Higgs has {\mbox{$ \Delta^{h^{0}}_{\alpha} =
  \Delta^{h^{0}}_{\alpha_{s}} = 0$} and {$ \Delta^{h^{0}}_{s^{2}} \ll
  \Delta^{top}_{s^{2}}$}}. We therefore neglect possible contributions
  to \del from the SM Higgs, and $\stho = 0.2324$ is consistent
  with $m_{h^{0}} = \mz$. We account then for different values of
  $m_{h^{0}}$  as a part of the $0.0003$ error bar.

  When evaluating \delt it is convenient to use
  $\sth = \stho = 0.2324$ rather than the
  one-loop prediction (as we could choose to do in case \aaa \cite{h1}).
  This induces a weak dependence of the \stho prediction on  the \stho
  input value via the first terms in (\ref{deltop1}) and (\ref{deltop2}),
  but in practice the effect is negligible. As a matter of fact, all the
  logarithmic contributions to \delt (and not just the ones which appear
  in $\Delta_{s^{2}}^{top}$) are negligible in comparison
  to the quadratic ones, and will be omitted later.
  Table 8 lists the different contributions to \delt (which are the
  same in the SM and the MSSM).

  Another issue that is related to the heavy top is the
  contribution of the top Yukawa coupling, \httns, to the two-loop
  \btf \cite{mv,fh,j4,j5}. If {\mbox{$\htt \approx 1$}}, we need to
  re-introduce the relevant term (that was neglected above)
  in the \btf (\ref{beta}), i.e.,
  \begin{equation}
  \mu{\frac{d\alpha_{i}}{d\mu}} = {\frac{\bi}{(2\pi)}}\alpha_{i}^{2}
  + \sum_{j = 1}^{3}{\frac{\bij}{(8\pi^{2})}}\alpha_{i}^{2}\alpha_{j}
  -\bitns{\frac{\httns^{2}}{(16\pi^{2})}}{\frac{\alpha_{i}^{2}}{(2\pi)}},
  \label {betay}
  \end{equation}
  where \bit can be calculated using, for example, ref. \cite{mv,j4}
  and are of the order of magnitude of unity. In the SM {\mbox{$\bit =
  {\frac{17}{10}},{\frac{3}{2}},2$} } for $i$ = 1, 2, 3, respectively
  \cite{mv,fh}.
  In the MSSM there are (to this order) two additional Yukawa
  terms in which a higgsino is coupled to a stop and a top. One then has
  \cite{j5} {\mbox{$\bit = {\frac{26}{5}},6,4$} } for $i$ = 1, 2, 3.
  \htt is running and is coupled to $\alpha_{i}$ at the one-loop order.
  \dely are functions of the couplings \htt and \alg at the unification
  point, and of the unification point parameter $t$, and have to be
  calculated numerically.

  Let us consider only the MSSM where the effect is relevant.
  A heavy top can then also
  imply a large Yukawa coupling for the $b$-quark, \hbbns, i.e.,
  {\mbox{$ \topbot = \tan\beta $} }  where {\mbox{$\tan\beta$} }
  is the ratio of the vacuum expectation values of the two Higgs doublets
  \cite{ikkt}. For a large enough {\mbox{$\tan\beta$} } one could have
  {\mbox{$\hbb \approx\htt$}} \cite{bbo}. However,
  such a situation is not very likely.
  Proton decay via dimension-five operators constrains $\tan\beta$
  (i.e., {\mbox{$ \tan\beta  \leq 4.7 $}  for  {$ \mt = 125$} GeV,
  assuming {$\als = 0.113 \pm 0.005$}}) \cite{nath}.
  We will keep neglecting
  \hbbns. (One should note that the requirement $\sin\beta < 1$ places
  a lower bound on \htt for a fixed \mtns.) We calculate the Yukawa
  correction by solving numerically the coupled RGE's \cite{j5}. The
  results are given in Table 9 in terms of the corrections to the
  predictions, $H_{s^{2}}$, $H_{\alpha_{s}}$, $H_{t}$, and $H_{\frac{1}
  {\alpha_{G}}}$, rather then in terms of \delyns.

  Instead of the full two-loop numerical calculation one could use
  an approximation in which \htt is constant.
  Then  the new term in (\ref{betay}) is realized
  as a negative correction to \bins, and
  \begin{equation}
  \dely \approx \bitns{\frac{\httns^{2}}{(16\pi^{2})}}t,
  \label{dely}
  \end{equation}
  or $\frac{\Delta_{i}^{Yukawa}}{h_{t}^{2}} \approx 0.17$, $0.20$,
  $0.13$, for $i$ = 1, 2, 3, respectively.
  One can see from Table 9
  that taking $\htt \approx 1 \approx h_{fixed}$
  is a reasonable approximation ($h_{fixed}$ is the fixed point of the
  one-loop top Yukawa RGE \cite{hill2,ibanez,j5,mc1}).

  Lastly, we consider contributions from non-renormalizable operators at
  the high-scale, which may be induced by the physics between \mg
  and \newline
  {\mbox{$\mpk \approx 1.22\times10^{19}$} GeV }\cite{hill,shaf}.
  We consider only dimension-five operators,
  {\mbox{$ -{\frac{1}{2}}{\frac{\eta}{M_{planck}}}$}
  Tr({$F_{\mu\nu}$}}{\mbox{$ \Phi F^{\mu\nu}$})},
  where \etaa is a dimensionless parameter and
  {\mbox{$F_{\mu\nu}$} } is the field strength tensor.
  In the \sug model {\mbox{$ \Phi$} } is the {\mbox{$\underline{24}$} }
  real-Higgs (Majorana super)multiplet.
  (Contributions from higher-dimension operators are suppressed by powers
  of \mppns.)  When $\Phi$ acquires an expectation value the effect is to
  renormalize the gauge fields, which can be absorbed into a redefinition
  of the couplings. It is shown in ref. \cite{hill,shaf} that the running
  couplings at \mg are related to the underlying gauge coupling
  $\alpha_{G}(\mg)$ by {\mbox{$ \aig = {\frac{(1 + \ei)}{\alg}} $}},
  where
  {\mbox{$ \ei = \eta\ki\sqrt{\frac{r}{\pi\alpha_{G}}}
  {\frac{M_{V}}{M_{planck}}}$}}.
  In the \sug model $r = \frac{2}{25}$ and
  {\mbox{$ \ki = {\frac{1}{2}},{\frac{3}{2}}, -1$} } for  $i$ =1, 2, 3,
  respectively.
  We treat these operators
  perturbatively (i.e., for {\mbox{$|\eta| < 10$}}), by defining
  \begin{equation}
  \deln \equiv -\eta\ki\sqrt{\frac{r}{\pi\alpha_{G}^{3}}}
                     {\frac{\mg}{\mpk}} \; ,
  \label{nro}
  \end{equation}
  where it is sufficient to use the one-loop expressions for \alg and
  {\mbox{$\mg = $}}{\mbox{$\mz e^{2\pi t}$}}.
  (\ref{nro}) is valid in the MSSM as well \cite{hill}, and different
  normalizations and scales can be absorbed in $\eta$.

  Like \thns, \deln and \dely depend on the input parameters
  through \ti and \algns. We use the full
  two-loop values for \ti and \alg when estimating these correction
  terms (consistent with solving for \th iteratively).
  At the price of a minor technical inconsistency, we always
  use the two-loop values of \ti and \alg given in Table 5a.

  The different contributions to \delns, in the SM and the MSSM,
  are listed in Tables 8 - 10.
  From Tables 10b and 8 we learn that different contributions to
  \del in the MSSM are a-priori comparable, and a comparision
  with Table 10a suggests that they are more significant, by number and
  magnitude, than in the SM. These points were stressed
  recently in ref. \cite{h2}.

  At this point one is able to write explicit expressions for \agns, $t$,
  and \stho (\agns, $t$ and \alsns).
  We give below those for \sthons, \alsns, $\ti(\alpha,s^{2}_{0})$ and
  $\alg(\alpha,s^{2}_{0})$
  in the MSSM, which are the main results of this section.
  We hereafter neglect all logarithmic contributions to \deltns.
  Constant correction terms are included in the functions $\delta_{i}$,
  which are normalized such that the conversion term is unity. Our best
  guesses for the values of the functions $H$ are $H_{s^{2}}= -0.0003^{
  +0.0002}_{-0.0001}$; $H_{\alpha_{s}} = -0.0010^{+0.0007}_{-0.0004}$;
   $H_{t} = -0.004^{+0.003}_{-0.002}$;
   $H_{\frac{1}{\alpha_{G}}} = +0.19^{+0.08}_{-0.14}$; corresponding to
   $\htt = 1$ at the unification point and the range given in Table 9.
  \begin{eqnarray}
  \stho  =  0.2 +
  {\frac{7}{15}}{\frac{\allp}{\als}}\left( 1 \pm \delta_{\alpha}
  \pm  \delta_{\alpha_{s}} \right)
  & &  \nonumber \\
  +   0.0031 +  H_{s^{2}} +
  {\frac{\allp }{60 \pi}}(\delta_{1} + \delta_{2}) & &
  \label{sin}
  \end{eqnarray}
  \begin{eqnarray}
  \als = {\frac{7 \allp}{15\stho - 3}}\left(1 \pm  \delta_{\alpha} \pm
  \delta_{s^{2}} \right)
  & & \nonumber \\
  + 0.012 +  H_{\alpha_{s}} +
  {\frac{28\allp^{2}} {{(60\stho - 12)}^{2}\pi}}(\delta_{1}
  + \delta_{2}) & &
  \label{alps}
  \end{eqnarray}
  \begin{eqnarray}
  t = {\frac{3 - 8\stho}{28\allp}}\left( 1 \pm \delta_{\alpha} \pm
  0.2\delta_{s^{2}} \right)
  & & \nonumber \\
  + 0.08 +  H_{t} +
  {\frac{5}{168\pi}}(\delta_{3} + \delta_{4} + \delta_{5})& &
  \label{titi}
  \end{eqnarray}
  \begin{eqnarray}
  \ag = {\frac{3 -36\stho}{28\allp}}\left( 1 \pm \delta_{\alpha} \pm
  0.2\delta_{s^{2}} \right)
  & & \nonumber \\
  -1.23  + H_{\ag}
  -{\frac{5}{168\pi}}(\delta_{3} + {\frac{33}{5}}\delta_{4}
  + \delta_{6}) & &
  \label{alphag}
  \end{eqnarray}
  where\footnote{Negligible inconsistencies between
  (\ref{del1})-(\ref{del6}) and Tables 8-10 may exist due to roundoff.}
  \begin{eqnarray}
  \delta_{1}  = 1 - 12\mvxa - 6\mvxb + 18\mvxc & & \nonumber \\
  + 25\miza - 100\mizb + 56\mizc
  & &
  \label{del1}
  \end{eqnarray}
  \begin{eqnarray}
  \delta_{2}  =
   + 3.9 + 47.4\left( (\mttt)^{2} - 1 \right) + 8.00\eta
   & &
   \label{del2}
   \end{eqnarray}
   \begin{eqnarray}
   \delta_{3}  =  -30\mvxa +{\frac{6}{5}}\mvxc
   + {\frac{15}{2}}\miza
   & &
   \label{del3}
   \end{eqnarray}
   \begin{eqnarray}
   \delta_{4}  =  1 + 18\mvxa -6\mvxb
   - {\frac{25}{2}}\mizb
   & &
   \label{del4}
   \end{eqnarray}
   \begin{eqnarray}
   \delta_{5}  =  + 7.6\left((\mttt)^{2} - 1\right) + 0.53\eta
   & &
   \label{del5}
   \end{eqnarray}
   \begin{eqnarray}
   \delta_{6}  = + 34.2\left((\mttt)^{2} - 1\right)  +5.01\eta
   & &
   \label{del6}
   \end{eqnarray}
   and
  \begin{eqnarray}
  \delta_{\alpha} = \allp\delta(\alp) & & \\
  \delta_{\alpha_{s}} = {\frac{\delta(\alsns)}{\als}} & & \\
  \delta_{s^{2}} = {\frac{\delta(\sthons)}{(\stho - 0.2)}} \; . & &
  \end{eqnarray}
  The third term in (\ref{sin}), and the
  second terms in (\ref{alps}), (\ref{titi}) and (\ref{alphag}) are
  two-loop terms.
  These, as well as $\delta_{2}$, $\delta_{5}$, $\delta_{6}$,
  and the functions $H$,
  depend weakly on the values of input parameters used.
  All the other expressions can be similarly constructed.
  Implications of these results are considered in the following
  section, where we also estimate the values of the correction terms and
  their uncertainties (see Table 11).

  %4444
  \newpage
  \section{ The Correction Terms in The MSSM}
  We are now equipped to discuss the correction terms in the MSSM
  (where their contribution is significant) more quantitatively.
  From (\ref{del1}) we can realize the meaning of
  the naive parameter \msus mentioned above, i.e.,
  \begin{equation}
  25\miza - 100\mizb + 56\mizc \equiv -19\msz.
  \label{msusy}
  \end{equation}
  That is, the effect of an arbitrary sparticle  spectrum on the \stho
  and \als predictions can always be parametrized in terms of the (same)
  parameter \msusns. On the other hand, the \ag and $t$ uncertainties
  have different dependences on the \miins.
  It is important to note that the coefficient on the $r.h.s.$
  of (\ref{msusy})
  is small due to cancellations, while those on the $l.h.s.$ are large.
  In the case {\mbox{$ M_{2} \gg M_{1}$}} and/or $M_{2} \gg M_{3}$
  mentioned above, the $l.h.s.$ of (\ref{msusy}) (and therefore
  $\delta_{1}$) can grow significantly, and \msus can then be large.
  However, excluding such a case implies that
  {\mbox{$ \msus \approx 1$} TeV }can be achieved only by some
  adjustment of the parameters. It is not enough to have large \mii in
  order to have a large \msusns. For example, ($\miia \approx \miib
  \approx 1$ TeV, $\miic \approx 2$ TeV) correspond to $\msus \approx
  130$ GeV  and ($\miia \approx 850$ GeV, $\miib \approx 840$ GeV,
  $\miic \approx 1$ TeV) correspond to $\msus \approx 495$ GeV.
  On the other hand, a small \msus does not imply a low spectrum.
  For example, ($ \miia \approx 550$ GeV, $\miib \approx 540$ GeV,
  $\miic \approx 980$ GeV) or ($ \miia \approx 600$ GeV, $\miib
  \approx \miic \approx 266$ GeV) both correspond to $\msus \approx \mz$.
  One can even have $\msus \ll \mz$ (a large positive contribution
  to $\delta_{1}$). For example,
  for the two spectra of ref. \cite{ross1} given
  in Table 7 we have {\mbox{$\msus \approx 32$} GeV and {$21$} GeV},
  respectively. Thus, \msus does not teach one about the actual
  spectrum, and the widely chosen range of $ \mz < \msus < 1$ TeV
  does not represent the possible sparticle spectra properly, as was
  emphasized in ref. \cite{ross1}.

  For $\mvvv = \mg$, as is suggested by proton decay
  constraints, the high-scale threshold contribution to
  {\mbox{$\delta_{1}$} } is always positive.
  This was emphasized recently in ref. \cite{ekn}.
  If one combines the two observations, a positive
  {\mbox{$\delta_{1}$} }is likely.
  Such a situation is not favored in the MSSM, as the predictions
  for \stho and \als are already slightly higher
  than the central input values of these parameters.
  (It could even signal the model failure if the \als
  value is determined to be near the lower end of the $0.120 \pm 0.010$
  range, as was also emphasized in ref. \cite{ekn}.) Requiring a
  negative $\delta_{1} - 1$ can then severely constrain the spectrum
  parameters. However, until the \mii are known in detail
  it is not clear to us that there is really a
  problem, and at the present time we do not find
  much point in  elaborating on the $\delta_{1} - 1$ sign.
  Furthermore, the above
  situation  can be compensated by a negative $\delta_{2}$, i.e.,
  if either {\mbox{$\eta < 0$} } or {\mbox{$\mt < 138$} GeV},
  or a combination of the two.  Theoretical knowledge
  of \deln is thus important for a more quantitative
  discusssion, especially once \als is more accurately known
  and the top is found.
  The discussion above stresses once again a major weakness
  of the MSSM --
  proton decay via dimension-five operators. If \mg is not strongly
  constrained, $\delta_{1} - 1$
  can be made negative without any constraints
  on the sparticle spectrum. This is the situation in a simple extension
  of the MSSM in which the discrete $Z_{2}$ R-parity
  is replaced by a discrete $Z_{3}$ baryon-parity,
  and the dimension-five operators that are responsible for the proton
  decay are forbidden \cite{ross2}.
  (Though the phenomenology of such a model is very different, it does
  not directly affect the discussion in this paper.)
  Similarly, superstring-derived models which are not true GUT's may not
  have any problems with proton decay.
  Finally, more (split) (super)multiplets at
  the high-scale boundary, within  \sug  or in a model with a larger
  GUT gauge group, can change the above situation as well.
  We discuss such a possibility in the following section.

  A similar discussion applies to $\delta_{3} + \delta_{4} + \delta_{5}$,
  but here the sparticle contribution can easily pick any sign; e.g,
  $\miib \geq \miia$ will give a negative contribution, and thus a lower
  \mgns. $\miia \gg \miib$ is thus favored by proton decay
  (which implies $\msus \ll 1$ TeV). One should also note that \ti
  will be  corrected for $\msus = \mz$.

  For a more quantitative discussion, one has to choose
  reasonable ranges for the different parameters. We suggest
  \begin{itemize}
  \item   {\mbox{$\mt \leq \mii \leq 1$} TeV},
  and further constrain the splitting to be less than a factor of 4.
  $\miib \gg \miia$ and/or $\miib \gg \miic$ are  excluded
  (and proton decay may exclude $\miic > \miia$).
  \item   {\mbox{$10^{-2}\times\mg \leq \mv,\mvv,\mvvv \leq \mg$} }
  and constrain \mvv and \mvvv to be smaller than a few times
  \mv (and proton decay may further constrain \mvvvns).
  \item   $0 \leq |\eta| \leq 10$.
  For larger values the treatment is not perturbative.
  Note that \deln becomes negligible for $|\eta| < 1$.
  For example, large-radius Calabi-Yau compactfication which yield
  interesting neutrino masses predict $|\eta| \ll 1$ \cite{pgl4}.
  \item   {\mbox{$113$} GeV {$ \leq \mt \leq 159$} GeV}
  from precision electroweak data.
  \end{itemize}
  For these ranges we present the different contributions to \sthons,
  \als -- ($\delta_{1} + \delta_{2}$) -- and to $t(\alpha,s^{2}_{0})$
  -- ($\delta_{3} + \delta_{4} +\delta_{5}$) -- in Figures 4-6. We
  also display in each figure the two-loop correction, the corresponding
  input error bar, and for \stho the prediction uncertainty
  from the \als input value error bar. (The \stho error bar induces much
  smaller uncertainties, and those induced by the \allp
  error bar are negligible.)

  The observations made above become clear if we examine once again
  the spectra given in ref. \cite{ross1}. The sparticle  and \mt
  contributions to $\delta_{1} + \delta_{2}$  can be offset by
  $\eta \approx -4.5$ and $-0.7$ for the two cases. Also the
  \mii and \mt contributions are comparable
  and in the second case come with opposite signs (which explains
  the small $|\eta|$ required in this case). If
  constraints from proton decay are ignored (see the footnote
  above), we can also use $\mvvv \approx 0.1\mg$ and $0.75\mg$.
  (The second
  value corresponds to {\mbox{$\mvvv \approx 2 \times 10^{16}$} GeV}.)
  Thus, we see that for a combination of
  $\msus < \mz $, $\mt < 138$ GeV,
  and a \mvvv just below the unification scale, we can still have
  $\delta_{1} + \delta_{2} \leq 1$.

  Finally, we estimate the theoretical uncertainties for the
  \sthons, \alsns, $\ti(\alpha,s^{2}_{0})$ and
  $\ag(\alpha,s^{2}_{0})$ predictions. We present these in Table 11.
  (The prediction for \stho is to be compared with the value in
  (\ref{input6}), for which
  \mt does not contribute to the error bar.)
  For our choices of values for the different correction parameters,
  we obtain theoretical uncertainties to \stho comparable with the
  one induced by the \als error bar, and in the \als case comparable
  with its error bar. They may add to or may offset each other. In order
  to have a more decisive observation, a better determination of \als
  and elimination of some of these uncertainties are required. We also
  obtain {\mbox{$\mg \geq 1.3 \times 10^{16}$} GeV
  where different corrections were added in quadrature.
  This is well above the limit ($\sim 10^{15}$ GeV) from
  proton decay via vector boson exchange \cite{proton}.
  Let us emphasize again that though we
  arbitrarily chose the different correction parameter values,
  our choices serve as reasonable order of magnitude estimations.

  %5555
  \newpage
  \section{A General Treatment of Threshold Correction at The
  High-Scale}
  Above, we assigned
  explicit mass parameters to the different (super)multiplet classes
  at the high-scale -- \mvns, \mvvns, \mvvv --  while
  at the low-scale boundary we parametrized the threshold corrections
  using three effective mass parameters -- \miians, \miibns, \miic --
  which can be computed in any model. Similar effective parameters
  can also be defined at the high-scale, i.e.,
  \begin{equation}
  \sum_{\zeta}{\frac{b_{i}^{\zeta}}{(2\pi)}}\mxg \equiv
  {\frac{b_{i}^{matter}}{(2\pi)}}\migi
  \; \forr
  \label{miprime}
  \end{equation}
  For definiteness we identify $\mv \equiv \mg$ where \mv here is
  the mass of the vector (super)fields, which we assume are degenerate.
  ( (\ref{miprime}) can be easily generalized to include non-degenerate
  vector masses.)  The summation
  is then over all massive matter -- scalar and fermion (Majorana,
  chiral, and Dirac) -- (super)fields at the high-scale boundary.
  $b^{matter}_{i}$ is the (decoupled)
  contribution of these (super)fields
  to \bins. By using the $M_{i}^{'}$, we lose some sensitivity
  to the fine details of the heavy spectrum,
  but are able to examine models in which there are
  more and larger supermultiplets. (We will limit ourselves, however,
  to consideration of simple
  extensions in which additional supermultiplets are
  decoupled at the high-scale boundary.)

  Assuming the heavy supermultiplets of the minimal (SUSY) \sug model,
  \begin{equation}
  \Delta_{1} = \frac{5}{4\pi}\miza +
  \frac{1}{5\pi}\miga + ...
  \label{delta1new}
  \end{equation}
  \begin{equation}
  \Delta_{2} = \frac{25}{12\pi}\mizb +
  \frac{1}{\pi}\migb +...
  \label{delta2new}
  \end{equation}
  \begin{equation}
  \Delta_{3} = \frac{2}{\pi}\mizc +
  \frac{2}{\pi}\migc + ... \; ,
  \label{delta3new}
  \end{equation}
   where we wrote explicitly
   only the \mii and \miip contributions. $\delta_{1}$ can
   then be rewritten as
   \begin{eqnarray}
   \delta_{1}  = 1 +4\miga - 48\migb + 56\migc & & \nonumber \\
   + 25\miza - 100\mizb + 56\mizc.
   & &
   \label{del1new}
   \end{eqnarray}
   We can further define a new effective parameter
   \mheavy (in analogy with \msusns)
   \begin{equation}
   \frac{\sum_{i=1}^{3}w_{i}\migi}{\sum_{i=1}^{3}w_{i}}
   \equiv \mhg,
   \label{mheavy}
   \end{equation}
   where $w_{i} = \frac{5}{2}
  \sum_{j,k=1}^{3}\frac{1}{2}\epsilon_{jki}(b_{j}-b_{k}) b_{i}^{matter}$,
   and where $\epsilon_{ijk}$ is the Levi-Civita symbol, and the factor
   of $\frac{5}{2}$ is introduced for consistency with (\ref{del1}).

   In the minimal (SUSY) \sug model this gives
   \begin{equation}
   4\miga - 48\migb + 56\migc \equiv 12\mhgns.
   \label{mheavy2}
   \end{equation}
   While \mizi are always positive, \migi can have either sign. If indeed
   $\mvvv \approx 3\mv$, then \miga is positive, \migb is more probably
   negative, and $\migc > \frac{3}{4}\migb$, which is a restatement
   of the high-scale threshold positive contribution to $\delta_{1}$
   discussed above. If we introduce more
   matter supermultiplets in (\ref{miprime}),
   this situation may change. Let us assume $n_{10}$ ($n_{5}$) additional
   $\underline{10}$ ($\underline{5}$)
   of \sug chiral supermultiplets\footnote{ Such a
   situation can arise in models in which all matter is embedded in
   $\underline{27}$ supermultiplets
   of $E_{6}$ at some scale $\mu$, $\mu \geq \mg$.
   Our assumptions imply that additional
   massive vector superfields are irrelevant, and that there will
   be no additional Majorana massive superfields. If the $E_{6}$
   model is derived from the string, then usually there are no adjoint
   representations, and therefore no Majorana supermultiplets.}.
   Each $\underline{10}$  (\underline{5}) consists of
   $(3,1,\frac{2}{3})\oplus(3,2,\frac{1}{6})\oplus(1,1,1)$
   ($(3,1,-\frac{1}{3})\oplus(1,2,\frac{1}{2})$)
   $S$ superfields, and we further
   allow an arbitrary split among the different $S$ thresholds
   introduced here.
   For illustration, we will also assume that the new superfields
   are not constrained by proton decay limits. In practice, the extent
   to which they are constrained is determined by their couplings
   to the MSSM superfields, and by discrete symmetries (e.g., R-parity)
   and their quantum number assignments.
   Then,  $\delta b_{i}^{matter} = \frac{3}{2}n_{10} + \frac{1}{2}n_{5}$
   for $i$ = 1, 2, 3,
   and
   \begin{eqnarray}
  (15n_{10}+5n_{5}+4)\miga - (36n_{10}+12n_{5}+48)\migb  & & \nonumber \\
   + (21n_{10}  + 7n_{5}+ 56)\migc \equiv 12\mhgns. & &
   \label{mheavy3}
   \end{eqnarray}
   $\mheavy < \mg$ is now possible if the split is such that
   $\miipa \, (\miipc) \ll \miipb,\mg $
   or $\miipa,\miipc < \miipb,\mg$.
   Note that now $\delta_{1}- 1$ can easily pick either a negative
   or a positive sign. For example, $\mheavy \approx 0.1\mg$ and
   $\msus \approx 0.25\mz$  would imply $\delta_{1} \approx 0$.

  %6666
  \newpage
  \section{Conclusions}
  In this paper we considered various correction terms.
  We introduced the effective parameters \mii (which sum the low-scale
  threshold corrections), realized the naive parameter \msus
  in terms of the \miins, and pointed out that \msus can differ
  significantly from the actual sparticle masses.
  We then introduced similar parameters, \miipns, at the high-scale,
  and a different and more explicit set of high-scale parameters
  when we considered the minimal \sug model, in which
  the colored triplet Higgs superfield
  threshold is strongly constrained. The parameters  \mii and \miip
  can be used to conveniently
  compare threshold correction terms in different models.

  The centeral predictions of the MSSM are slightly high, but lie well
  within the experimental error bars. $Z$-pole determinations of \als
  favor no correction or a positive correction to the \als
  prediction, while low-energy determinations favor a negative
  correction. However, we showed that the magnitude and sign of the
  corrections to the two-loop predictions are determined by an interplay
  among various comparable terms. Of these terms, only one has a fixed
  sign: the contribution from the high-scale thresholds in the minimal
  SUSY-\sug model is positive when proton decay constraints are imposed.
  We pointed out that once simple extensions are considered, i.e., more
  heavy supermultiplets or replacing R-parity with baryon-parity, the
  above sign is no longer fixed. The sparticle contribution can be either
  positive (as for the two spectra of ref. \cite{ross1}) or negative,
  and so are the contributions from \mt and NRO's. Therfore we concluded
  that elaboration on the sign of any of these correction terms cannot
  be well justified at the present.

  The MSSM  then agrees well with experiment, and
  a theoretical uncertainty of $\sim + 0.0026 -0.0023$ ($+0.010 -0.008$)
  has to be assigned to the \stho (\alsns) prediction of the model.
  This is not the case when the SM is considered. Neither
  perturbative correction terms nor additional Higgs doublets can
  reverse the failure of coupling constant unification in this model.
  For example, the equivalent theoretical uncertainties in the SM
  are roughly $\sim \pm 0.0007$ and $\sim \pm 0.001$, respectively.
  The correction terms discussed, though
  negligible in the SM, may play an important role in the MSSM once
  more precise data is available.

  Finally, we would like to mention once again that
  we have used an alternative definition of the
  \ms weak angle \cite{s1}, which slightly differs from the canonical
  one in the way the $t$-quark is treated.

  %7777
  \section*{Acknowledgments}
  This work was supported by the Department of Energy Grant
  DE-AC02-76-ERO-3071. It is pleasure to thank Mirjam Cveti{\chc} and
  Alberto Sirlin for useful discussions.

%bbbb

 %tttt
 \newpage
 \section*{Table Captions}

  Table 1: Values of \alsns, adapted from \cite{alphas}. $R_{\tau}$
  refers to the ratio of hadronic to leptonic $\tau$ decays; $DIS$ to
  deep-inelastic scattering; $\Upsilon$, $J/\Psi$ to onium decays;
  and LEP($R$) to the ratio of hadronic to leptonic $Z$ decays.
  LEP(events) refers to the event topology in $Z \longrightarrow$ jets.
  This value was derived using resummed QCD \cite{alphas}, in which both
  $\alpha_{s}^{2}$ and next-to-leading logarithms are used in
  theoretical expressions (the same data would yield $0.119 \pm 0.006$
  using the $\alpha_{s}^{2}$ expressions). We choose $\als = 0.12
  \pm 0.01$ as a reasonable estimate of the average.

  Table 2a: $t$, \ag and \sth (or alternatively \sthons)
  predictions in terms  of \allpns, \als
  and the \btf coefficients, case \aaans.
  The correction terms are of the same form as the two-loop terms,
  only with \th replaced by \minus\delns.

  Table 2b: $t$, \ag and \als predictions in terms  of \allpns, \sth
  (or alternatively \sthons)
  and the \btf coefficients, case \bbbns.
  The correction terms are of the same form as the two-loop terms,
  only with \th replaced by \minus\delns.

 Table 3: The \btf coefficients \cite{j2}
 and their linear combination \newline \ddd.

 Table 4a: Two-loop terms for the case \aaa calculated using
 one loop values for the parameters (OL), and iteratively (TL).

 Table 4b: Two-loop terms for the case \bbb calculated using
 one loop values for the parameters (OL), and iteratively (TL).

 Table 5a: Numerical predictions of \tins, \agns, and \stho
 in  case \aaans.
 Input parameters are indicated by brackets.
 No correction terms are included.

 Table 5b: Numerical predictions of \tins, \agns, and \als
 in  case \bbbns.
 Input parameters are indicated by brackets.
 No correction terms are included.
 The near equality of cases \aaa and \bbb for the MSSM is a reflection
 of the success of the coupling constant unification.

 Table 6a: The two-loop
 predictions of the SM and MSSM in case \aaa
 for $\Delta F = 1$ additional fermion family and $\Delta n_{H} = 1$
 or 2 additional light Higgs (super)multiplets.
 Input parameters are indicated by brackets.
 No correction terms are included.

 Table 6b: The two-loop
 predictions of the SM and MSSM in case \bbb
 for $\Delta F = 1$ additional fermion family and $\Delta n_{H} = 1$
 or 2 additional light Higgs (super)multiplets.
 Input parameters are indicated by brackets.
 No correction terms are included.
 Note that for a negative \als the Taylor expansion is not valid.

 Table 7: The MSSM low-energy parameters calculated for the
 spectra of ref. \cite{ross1}.
 An \sul doublet is identified with its heavier member. Masses are in
 GeV.

 Table 8: The top correction terms \deltns.
 (These are common for the SM and the MSSM.)

 Table 9: The corrections to the predictions in the MSSM
     due to different values of the top Yukawa coupling, \httns,
     at the unification point. $\tan\beta$ is calculated
     using $\mt = 138$ GeV, and is not required to obey any limits.
     $\sin \beta < 1$ gives a lower bound on \httns. The corrections
     are denoted by $H$, with self-explanatory subscripts.

 Table 10a: The different correction terms \del in the
    (minimal \sugns) SM.

 Table 10b: The different correction terms \del in the
   (minimal \sugns) MSSM.

 Table 11: The different contributions to the theoretical uncertainties
 of the \sthons, \alsns, $t(\alpha,s^{2}_{0})$ and
 $\ag(\alpha,s^{2}_{0})$ predictions in the MSSM.
 The ranges of the parameters and the corresponding
 uncertainties serve as an order of magnitude estimate only, and in some
 cases are chosen to be smaller than those displayed in the figures.

%ffff
\newpage
\section*{Figure Captions}

  Figure 1: Predictions for \als from \allp and \sth in ordinary (SM)
  and SUSY (MSSM) GUT's. In the SM case the uncertainty $\sim \pm 0.001$
  includes that from \allp and \sth and the negligible high-scale and
  NRO errors. For the MSSM the small error bars are from \allp and \sth
  (including the \mt dependence) for $\msus =  \mz$ and $\msus = 1$ TeV.
  (We discuss the choice $\msus = 1$ TeV in section 4.)  The larger error
  bar includes the SUSY, high-scale, and NRO
  uncertainties added in quadrature. Various experimental determinations
  along with their nominal uncertainties are also shown.
  The dashed lines are the range $0.12 \pm 0.01$.

  Figure 2: Predictions for \sth from \allp and \als in ordinary (SM)
  and SUSY (MSSM) GUT's, compared with the region
  allowed by the data at $90\%$ C.L.
  The smaller ranges of uncertainties are from \als and \allp
  only, while the larger ones include the various low and high scale
  uncertainties added in quadrature. The predictions for $\msus = \mz$
  and $1$ TeV (see discussion in section 4) are shown for comparison.

  Figure 3: The running coupling in ordinary (SM) and SUSY (MSSM)
  GUT's, assuming $\sth = 0.2324 \pm 0.0006$, $\alp = 127.9 \pm 0.2$
  (the larger uncertainty compared to (\ref{input3}) is  due to \mtns),
  and $\als = 0.120 \pm 0.010$, for $\msus = \mz$. The uncertainties
  from threshold effects are best seen in Figures 1 and 2.

  Figure 4a - 4d: Contributions of individual correction terms
 --  the SUSY effective mass parameters $M_{i}$ (4a);
 the heavy thresholds at the high scale (4b); the top (4c); and NRO's
 at the high scale (4d) -- to the \stho prediction
 (via the last term of (\ref{sin})).
 The NRO term changes sign for $\eta < 0$.
 The error bar on \stho (dashed line),
 the uncertainty induced by the error bar on \als (dash-dot line),
 and the two-loop contribution to the \stho prediction
 (dotted line) are given for comparison.

 Figure 5a - 5d: Contributions of individual correction terms
 --  the SUSY effective mass parameters $M_{i}$ (5a);
 the heavy thresholds at the high scale (5b); the top (5c); and NRO's
 at the high scale (5d) --
 to the \als prediction (via the last term of (\ref{alps})).
 The error bar on \als (dashed line) and
 the two-loop contribution to the \als prediction (dotted line) are given
 for comparison.

 Figure 6a - 6d: Contributions of individual correction terms
 --  the SUSY effective mass parameters $M_{i}$ (6a);
 the heavy thresholds at the high scale (6b); the top (6c); and NRO's
 at the high scale (6d) --
 to the prediction of the scale parameter $t$
 (via the last term of (\ref{titi})).
 The two-loop contribution to $t$ (dotted line) is given
 for comparison.

 %tttt
 \newpage
 \section*{ }

\oddsidemargin -0.1in

\begin{tabular}{|c c|} \hline
 source  & \als  \\ \hline \hline
 $R_{\tau}$
 &$0.118 \pm 0.005$
 \\ \hline
 $DIS$
 & $0.112 \pm 0.005$
 \\ \hline
 $\Upsilon$, $J/\Psi$
 & $0.113 \pm 0.006$
 \\ \hline
 LEP($R$)
 & $0.133 \pm 0.012$
 \\ \hline
 LEP(events)
 & $0.123 \pm 0.005$
 \\ \hline
 average
 & $0.120 \pm 0.010$
 \\ \hline
 \multicolumn{2}{c}{Table 1} \\
 \end{tabular}
 \vspace{0.2in}

 \begin{tabular}{|c|c|c|} \hline
 & one-loop term &  two-loop term  \\  \hline \hline
 $t$ & {\mbox{$ {\frac{1}{D}}\left( {{\frac{3}{\alpha(M_{Z})}}
 -{\frac{8}{\alpha_{s}(M_{Z})}} }  \right)  $}}
 & {\mbox{$ -{\frac{1}{D}}( 5\tha + 3\thb - 8\thc )$}}
 \\  \hline \hline
 \ag  & {\mbox{$ {\frac{1}{D}} \left(  {\frac{-3b_{3}}{\alpha(M_{Z})}} +
   {\frac{5b_{1} + 3b_{2}}{\alpha_{s}(M_{Z})}}   \right)  $}}
 & {\mbox{$ -{\frac{1}{D}}\left({ (5\ba + 3\bb)\thc - \bc(5\tha + 3\thb)}
      \right) $}}
  \\  \hline \hline
 \sth  & {\mbox{$ {\frac{1}{D}}\left( { 3(\bb - \bc) + 5(\ba - \bb)
 {\frac{\alpha(M_{Z})}{\alpha_{s}(M_{Z})}}} \right) $}}
 & {\mbox{$ -{\frac{5\alpha(M_{Z})}{D}}\left({( \bb - \bc)\tha
 + (\bc - \ba)\thb
 + (\ba - \bb)\thc} \right) $}}
 \\  \hline
 \multicolumn{3}{c}{Table 2a} \\
 \end{tabular}
 \vspace{0.2in}

 \begin{tabular}{|c|c|c|} \hline
 & one-loop term &  two-loop term   \\  \hline \hline
 $t$ & {\mbox{${\frac{3 - 8s^{2}(M_{Z})}{5(b_{1}- b_{2})
          \alpha(M_{Z})}} $}}
 & {\mbox{$ +{\frac {\theta_{2} - \theta_{1}}{b_{1} - b_{2}}} $}}
 \\  \hline  \hline
 \ag  & {\mbox{$ {\frac {3b_{2}(1-s^{2}(M_{Z})) - 5b_{1}s^{2}(M_{Z})}
  {5(b_{2}-b_{1}) \alpha(M_{Z})}}  $}}
 & {\mbox{$ +{\frac { b_{2}\theta_{1} - b_{1}\theta_{2} }
  {b_{1} - b_{2}} } $}}
  \\ \hline  \hline
 \als  & {\mbox{${\frac{5(b_{1}-b_{2})\alpha(M_{Z})}
            {Ds^{2}(M_{z}) - 3(b_{2}-b_{3})}} $}}
 & {\mbox{$ -{\frac{25(b_{1} - b_{2})\alpha(M_{Z})^{2}}
 {\left( Ds^{2}(M_{Z}) - 3(b_{2} - b_{3}) \right)^{2}} }
 \left({( \bb - \bc)\tha + (\bc - \ba)\thb + (\ba - \bb)\thc} \right) $}}
 \\ \hline
 \multicolumn{3}{c}{Table 2b} \\
 \end{tabular}
 \newpage

 \begin{tabular}{|c|c|c|} \hline
 & SM & MSSM \\                                \hline
 \bi & {\mbox{$ \left( \begin{array}{c} \ova  \\
 \ovb   \\ -7 \end{array} \right) $}}
 & {\mbox{$ \left( \begin{array}{c} \ovc \\ 1
   \\
   -3 \end{array} \right) $}} \\               \hline
  \bij & {\mbox{$ \left( \begin{array}{c c c} 3.98 & 2.7 & 8.8 \\
       0.9 & {\mbox{$\frac{35}{6}$}} & 12  \\
       1.1 & 4.5 & -26   \end{array} \right)  $}}
   & {\mbox{$ \left( \begin{array}{c c c} 7.96 & 5.4 & 17.6 \\
       1.8 & 25 & 24  \\
       2.2 & 9 & 14   \end{array} \right)  $}} \\  \hline
  $D$ &  67 & 60 \\ \hline
 \multicolumn{3}{c}{Table 3} \\
 \end{tabular}
 \vspace{0.2in}

 \begin{tabular}{|c|c|c|c|c|}  \hline
 two-loop term & \multicolumn{2}{c|}{SM} &
   \multicolumn{2}{c|}{MSSM}  \\ \cline{2-5}
      & OL & TL & OL &TL \\ \hline
 \tha & 0.22 & 0.21 & 0.67 & 0.69 \\  \hline
 \thb & 0.29 & 0.28 & 1.09 & 1.13 \\  \hline
 \thc & $-0.41$ & $-0.40$ & 0.56 & 0.58 \\  \hline
 \multicolumn{5}{c}{Table 4a} \\
\end{tabular}
\vspace{0.2in}

\begin{tabular}{|c|c|c|c|c|}  \hline
 two-loop term & \multicolumn{2}{c|}{SM} &
   \multicolumn{2}{c|}{MSSM}  \\ \cline{2-5}
      & OL & TL & OL &TL \\ \hline
 \tha & 0.16 & 0.16 & 0.64 & 0.71 \\  \hline
 \thb & 0.21 & 0.21 & 1.05 & 1.16 \\  \hline
 \thc & $-0.27$  & $-0.28$ & 0.54 & 0.60\\  \hline
 \multicolumn{5}{c}{Table 4b} \\
\end{tabular}
\newpage

\begin{tabular}{|c|c|c||c|c|} \hline
 & \multicolumn{2}{c||}{SM} & \multicolumn{2}{c|}{MSSM}
 \\ \cline{2-5}
 & one-loop & two-loop& one-loop & two-loop
 \\ \hline \hline
 \ti
 & 4.73
 & 4.65
 & 5.28
 & 5.25
 \\ \hline
 \ag
 & 41.46
 & 41.32
 & 24.18
 & 23.49
 \\ \hline \hline
 \alp
 & \multicolumn{4}{c|}{(127.9)}
 \\ \hline
 \stho
 & 0.2070
 & 0.2100
 & 0.2304
 & 0.2335
 \\ \hline
 \als
 & \multicolumn{4}{c|}{(0.120)}
 \\ \hline
 \multicolumn{5}{c}{Table 5a} \\
\end{tabular}
\vspace{1.0in}

\begin{tabular}{|c|c|c||c|c|} \hline
 & \multicolumn{2}{c||}{SM} & \multicolumn{2}{c|}{MSSM}
 \\ \cline{2-5}
 & one-loop & two-loop& one-loop & two-loop
 \\ \hline \hline
 \ti
 & 4.01
 & 4.02
 & 5.21
 & 5.29
 \\ \hline
 \ag
 & 42.44
 & 42.25
 & 24.51
 & 23.28
 \\ \hline \hline
 \alp
 & \multicolumn{4}{c|}{(127.9)}
 \\ \hline
 \stho
 & \multicolumn{4}{c|}{(0.2324)}
 \\ \hline
 \als
 & 0.070
 & 0.072
 & 0.113
 & 0.125
 \\ \hline
 \multicolumn{5}{c}{Table 5b} \\
 \end{tabular}
 \vspace{0.2in}
 \newpage

\begin{tabular}{|c|c|c||c|c|} \hline
 & \multicolumn{2}{c||}{SM} & \multicolumn{2}{c|}{MSSM}
 \\ \cline{2-5}
 &$\Delta F=1$ & $\Delta n_{H}=1$ & $\Delta F = 1$ & $\Delta n_{H} = 2$
 \\ \hline \hline
 \ti
 & 4.69
 & 4.58
 & 5.32
 & 4.76
 \\ \hline
 \ag
 & 34.83
 & 40.83
 & 11.52
 & 22.09
 \\ \hline \hline
 \alp
 & \multicolumn{4}{c|}{(127.9)}
 \\ \hline
 \stho
 & 0.2099
 & 0.2141
 & 0.2345
 & 0.2562
 \\ \hline
 \als
 & \multicolumn{4}{c|}{(0.120)}
 \\ \hline
 \multicolumn{5}{c}{Table 6a} \\
\end{tabular}
\vspace{0.2in}

\begin{tabular}{|c|c|c||c|c|} \hline
 & \multicolumn{2}{c||}{SM} & \multicolumn{2}{c|}{MSSM}
 \\ \cline{2-5}
 &$\Delta F=1$ & $\Delta n_{H}=1$ & $\Delta F = 1$ & $\Delta n_{H} = 2$
 \\ \hline \hline
 \ti
 & 4.05
 & 4.06
 & 5.41
 & 5.78
 \\ \hline
 \ag
 & 36.70
 & 41.67
 & 10.69
 & 15.82
 \\ \hline \hline
 \alp
 & \multicolumn{4}{c|}{(127.9)}
 \\ \hline
 \stho
 & \multicolumn{4}{c|}{(0.2324)}
 \\ \hline
 \als
 & 0.072
 & 0.077
 & 0.130
 & negative
 \\  \hline
 \multicolumn{5}{c}{Table 6b} \\
 \end{tabular}
 \vspace{0.2in}

 \begin{tabular}{|c|c|c|c|c||c|c|c|c|} \hline
     \multicolumn{5}{|c||}{High Scale Parameters }
    & \multicolumn{4}{c|}{Low Scale Parameters }
    \\ \hline
   \mgg & \mss & \mmx & $A$ & $B$ & \mt &\mia & \mib & \mic \\ \hline
   140 & 190 & 190  & 0 & 0 & 160 & 261 & 207 & 352 \\ \hline
   230 & 120 &$-120$ & 0 & 0 & 100 & 282 & 245 & 527 \\ \hline
  \multicolumn{9}{c}{Table 7} \\
  \end{tabular}
  \newpage

\begin{tabular}{|c|c|c|c|} \hline
 & constant  & logarithmic & quadratic
 \\
 & term  & term & term
  \\ \hline \hline
 \dela
 & {\mbox{$ -0.15 $}}
 & {\mbox{$ +0.13\ln(\mttt) $}}
 & {\mbox{$ +0.15(\mttt)^{2}$}}
 \\ \hline \hline
 \delb
 & {\mbox{$ +0.25 $}}
 & {\mbox{$ +0.065\ln(\mttt) $}}
 & {\mbox{$ -0.25(\mttt)^{2}$}}
 \\ \hline \hline
 \deld
 & {\mbox{$ +0.04 $}}
 & {\mbox{$ +0.105\ln(\mttt) $}}
 & {\mbox{$ \minus$}}
 \\ \hline
 \multicolumn{4}{c}{Table 8} \\
\end{tabular}
\vspace{0.2in}

\begin{tabular}{|c|c|c||c|c|c||c|c|c|} \hline
 & & & \multicolumn{3}{c||}{case \aaa} & \multicolumn{3}{c|}{case \bbb}
 \\ \cline{4-9}
  $h_{t}(M_{G})$ &$h_{t}(M_{Z})$& $\tan\beta$
 & $H_{s^{2}}$ & $H_{t}$ & $H_{\frac{1}{\alpha_{G}}}$
 & $H_{\alpha_{s}}$ & $H_{t}$ & $H_{\frac{1}{\alpha_{G}}}$
 \\ \hline \hline
 $0.300$
 & $0.794$
 & $17.13$
 &$-0.00008$
 & $+0.002$
 & $+0.04$
 & $-0.0003$
 & $-0.001$
 & $+0.05$ \\  \hline
 $0.400$
 & $0.903$
 & $1.84$
 &$-0.00012$
 & $+0.003$
 & $+0.06$
 & $-0.0004$
 & $-0.002$
 & $+0.08$ \\  \hline
 $0.600$
 & $1.015$
 & $1.25$
 &$-0.00019$
 & $+0.004$
 & $+0.09$
 & $-0.0006$
 & $-0.003$
 & $+0.12$ \\  \hline
 $0.800$
 & $1.067$
 & $1.11$
 &$-0.00024$
 & $+0.005$
 & $+0.12$
 & $-0.0008$
 & $-0.004$
 & $+0.16$ \\  \hline
 $1.000$
 & $1.095$
 & $1.05$
 &$-0.00029$
 & $+0.006$
 & $+0.15$
 & $-0.0010$
 & $-0.004$
 & $+0.19$ \\  \hline
 $1.200$
 & $1.111$
 & $1.02$
 &$-0.00033$
 & $+0.007$
 & $+0.17$
 & $-0.0012$
 & $-0.005$
 & $+0.22$ \\  \hline
 $1.400$
 & $1.122$
 & $1.00$
 &$-0.00037$
 & $+0.008$
 & $+0.18$
 & $-0.0013$
 & $-0.005$
 & $+0.25$ \\  \hline
 $1.600$
 & $1.129$
 & $0.98$
 &$-0.00040$
 & $+0.009$
 & $+0.20$
 & $-0.0014$
 & $-0.006$
 & $+0.27$ \\  \hline
 \multicolumn{9}{c}{Table 9} \\
\end{tabular}
\vspace{0.2in}

 \begin{tabular}{|c|c|c|c|c|c|} \hline
 & \delc  & \delv & \delvv & \delvvv & \deln
  \\ \hline \hline
 \dela
 & \minus
 & {\mbox{$ -{\frac{35}{4\pi}} \mvxa $}}
 & \minus
 & {\mbox{$ +{\frac{1}{30\pi}} \mvxc $}}
 & {\mbox{$-0.0008\eta $}}
 \\ \hline \hline
 \delb
 & {\mbox{$ -{\frac{1}{6\pi}}  $}}
 & {\mbox{$ -{\frac{21}{4\pi}} \mvxa $}}
 & {\mbox{$ +{\frac{1}{6\pi}} \mvxb $}}
 & \minus
 & {\mbox{$-0.0024\eta $}}
 \\ \hline \hline
 \deld
 & {\mbox{$ -{\frac{1}{4\pi}}  $}}
 & {\mbox{$ -{\frac{7}{2\pi}} \mvxa $}}
 & {\mbox{$ +{\frac{1}{4\pi}} \mvxb $}}
 & {\mbox{$ +{\frac{1}{12\pi}} \mvxc $}}
 & {\mbox{$+0.0016\eta $}}
 \\ \hline
 \multicolumn{6}{c}{Table 10a} \\
\end{tabular}

\begin{tabular}{|c|c|c|c|c|c|c|} \hline
 & \delc  & \delv & \delvv & \delvvv & \dels & \deln
 \\ \hline \hline
 \dela
 & \minus
 & {\mbox{$ -{\frac{5}{\pi}} \mvxa $}}
 & \minus
 & {\mbox{$ +{\frac{1}{5\pi}} \mvxc $}}
 & {\mbox{$ +{\frac{5}{4\pi}} \miza $}}
 & {\mbox{$-0.014\eta $}}
 \\ \hline \hline
 \delb
 & {\mbox{$ -{\frac{1}{6\pi}}  $}}
 & {\mbox{$ -{\frac{3}{\pi}} \mvxa $}}
 & {\mbox{$ +{\frac{1}{\pi}} \mvxb $}}
 & \minus
 & {\mbox{$ +{\frac{25}{12\pi}} \mizb $}}
 & {\mbox{$-0.042\eta $}}
 \\ \hline \hline
 \deld
 & {\mbox{$ -{\frac{1}{4\pi}}  $}}
 & {\mbox{$ -{\frac{2}{\pi}} \mvxa $}}
 & {\mbox{$ +{\frac{3}{2\pi}} \mvxb $}}
 & {\mbox{$ +{\frac{1}{2\pi}} \mvxc $}}
 & {\mbox{$ +{\frac{2}{\pi}} \mizc $}}
 & {\mbox{$+0.028\eta $}}
 \\ \hline
 \multicolumn{7}{c}{Table 10b} \\
\end{tabular}
\vspace{0.2in}

\begin{tabular}{|c|c|c|c|c|} \hline
 & \stho & \als & $t(\alpha,s^{2}_{0})$ & $\ag(\alpha,s^{2}_{0})$
 \\ \hline \hline
 input value
 & $0.2324$
 & $0.120$
 & \minus
 & \minus
 \\ \hline
 error bar
 & $\pm 0.0003$
 & $\pm 0.010$
 & \minus
 & \minus
 \\ \hline \hline
 one-loop prediction
 & $0.2304$
 & $0.113$
 & $5.21$
 & $24.51$
 \\ \hline
 two-loop correction
 & $+0.0031$
 & $+0.012$
 & $+0.08$
 & $-1.23$
 \\ \hline
 Yukawa correction ($H$)
 & $-0.0003$
 & $-0.001$
 & $-0.004$
 & $+0.19$
 \\ \hline
 constant correction
 & $+0.0002$
 & $+0.001$
 & $+0.01$
 & $-0.06$
 \\ \hline \hline
 \als error bar
 & $\pm 0.0025$
 & \minus
 & \minus
 & \minus
 \\ \hline
 \stho error bar
 & \minus
 & $\pm 0.001$
 & $\pm 0.01$
 & $\mp 0.04$
 \\ \hline \hline
 $\miia = 4\mz, \miib = \miic = \mz$
 & $+0.0014$
 & $+0.005$
 & $+0.10$
 & $-0.10$
 \\ \hline
 $\miia = \miib = \miic = 6\mz$
 & $-0.0014$
 & $-0.005$
 & $-0.08$
 & $+1.27$
 \\ \hline \hline
 $\mt = 159$ GeV
 & $+0.0006$
 & $+0.002$
 & $+0.02$
 & $-0.10$
 \\ \hline
 $\mt = 113$ GeV
 & $-0.0006$
 & $-0.002$
 & $-0.02$
 & $+0.10$
 \\ \hline \hline
 $\mv = 0.3\mgns, \mvv = 0.05\mgns, \mvvv = \mg$
 & $+0.0013$
 & $+0.005$
 & $+0.31$
 & $-0.11$
 \\ \hline
 $\mv =  \mvv = \mgns, \mvvv = 0.5\mg$
 & $-0.0005$
 & $-0.002$
 & $-0.01$
 & $+0.01$
 \\ \hline \hline
 $\eta = \pm 5$
 & $\pm 0.0016$
 & $\pm 0.006$
 & $\pm 0.025$
 & $\mp 0.23$
 \\ \hline
 \multicolumn{5}{c}{Table 11} \\
\end{tabular}

\end{document}